\documentclass[12pt,preprint]{aastex}

\newcommand{\ltsima}{$\; \buildrel < \over \sim \;$}
\newcommand{\simlt}{\lower.5ex\hbox{\ltsima}} 
\newcommand{\gtsima}{$\; \buildrel > \over \sim \;$}
\newcommand{\simgt}{\lower.5ex\hbox{\gtsima}} 

\newcommand{\xmm}{{\emph{XMM-Newton}}}

\newcommand{\asca}{{\emph{ASCA} }}
\newcommand{\lum}{erg~s$^{-1}$}
\newcommand{\flux}{{erg~cm$^{-2}$~s$^{-1}$ }}

\newcommand{\sorg}{MCG--5-23-16}
\newcommand{\chandra}{{\emph{Chandra}}}

\received{}
\begin{document}

\title{Relativistic Iron K Emission and absorption  in the Seyfert 1.9  galaxy MCG--5-23-16}

\author{V. Braito\altaffilmark{1,2} , J.N.
Reeves\altaffilmark{1,2,3}, G.C. Dewangan\altaffilmark{4}, I. George\altaffilmark{1,5}, R.E. Griffiths\altaffilmark{4}, A.
Markowitz\altaffilmark{1,2},  K. Nandra\altaffilmark{6}, D. Porquet\altaffilmark{7},  A. Ptak\altaffilmark{1,2}, T.J.
Turner\altaffilmark{1,5}, T. Yaqoob\altaffilmark{1,2}, K. Weaver\altaffilmark{1} 
\altaffiltext{1}{Astrophysics Science Division, Code 662, NASA/Goddard Space Flight Center, Greenbelt, MD 20771, USA; vale@milkyway.gsfc.nasa.gov}
\altaffiltext{2}{Department of Physics and Astronomy, Johns Hopkins University, Baltimore, MD 21218.}
\altaffiltext{3}{Astrophysics Group, School of Physical and Geographical Sciences, Keele University, Keele, Staffordshire ST5 5BG}
\altaffiltext{4}{Department of Physics, Carnegie Mellon University, 5000 Forbes Avenue, Pittsburgh, PA 15213}
\altaffiltext{5}{Department of Physics, University of Maryland Baltimore County, 1000 Hilltop Circle, Baltimore, MD 21250, USA }
\altaffiltext{6}{Astrophysics Group, Imperial College London, Blackett Laboratory, Prince Consort Road, London SW7 2AW, UK }
\altaffiltext{7}{UMR 7550 du CNRS, Observatoire de Strasbourg,  11 rue de l'Universite, 67000 Strasbourg, France}
}

\begin{abstract}

We present the results of the simultaneous  deep \xmm\ and  \emph{Chandra} 
observations of  the bright Seyfert 1.9  galaxy MCG$-$5-23-16, which is thought to have one
of the best known examples of a relativistically broadened iron K$\alpha$ line.    The time
averaged spectral analysis shows that the  iron K-shell complex  is   best modeled with an
unresolved  narrow  emission component (FWHM $ < 5000$ km/s, EW $\sim 60 $ eV) plus a  broad
component.  This latter component has  FWHM $\sim 44000$ km/s  and EW $\sim $ 50 eV. Its
profile is well  described by an emission line   originating from an accretion disk 
viewed with an inclination angle $\sim 40^\circ$ and  with the emission  arising from 
within a few tens of gravitational radii of  the central black hole.  The time-resolved
spectral analysis of the \xmm\  EPIC-pn spectrum  shows  that both the narrow and  broad
components of   the Fe K emission line appear to  be constant in time within the errors. We
detected a narrow   sporadic absorption line at 7.7 keV  which appears to be variable on a
time-scale  of 20 ksec. If  associated with  {Fe\,\textsc{xxvi}} Ly$\alpha$ this absorption
is indicative of a possibly  variable, high ionization, high velocity outflow. The
variability of this absorption feature appears to rule out a local ($z=0$) origin. The
analysis of the  \xmm\ RGS spectrum   reveals   that the soft X-ray emission of
MCG--5-23-16 is likely  dominated by several emission lines superimposed on an unabsorbed 
scattered power-law continuum. The lack of  strong Fe L shell emission together  with the
detection of a strong forbidden line in the  {O\,\textsc{vii}}  triplet   is consistent
with a scenario where the soft  X-ray emission  lines are produced in a   plasma 
photoionized by the nuclear emission.\\

\end{abstract}


\section{Introduction} 

One of the   key  issues in  high energy research  on  Active
Galactic Nuclei (AGN) is the study   of the  6.4 keV  iron K$\alpha$ emission line 
profile,  which can provide  fundamental diagnostics of the physical  and dynamical
conditions of AGN central engines.   The fluorescent Fe K$\alpha$  emission line is a
prominent and  ubiquitous feature in the X-ray spectra  of AGN and it is   believed to be 
produced in the innermost regions of the AGN, such as  the Broad Line Region (BLR),  the 
circumnuclear obscuring torus and/or the accretion disk.  The profile  of the line itself
provides direct information on the  region from which it originates. If the Fe K$\alpha$
emission line is produced  far from the nucleus, e.g.  in the putative torus, its profile is
expected to be narrow, while  if the  line originates in the innermost part of the accretion
flow  a broad and   asymmetric  profile is predicted as a result of the special and general
relativistic effects  such as Doppler shifts, gravitational redshift and light bending (see
\citealp{Fabian2000,Reynolds03} for a review). In the latter case,  the shape of the profile
itself could be used to  derive information on the nature of the black hole and  accretion
disk system. 

 The observations with \asca\  ({\it Advanced Satellite for Cosmology and Astrophysics}) of relativistically    broadened iron Fe K$\alpha$ emission lines in
AGN  \citep{Nandra97}, and in particular the detection of a  broad and  skewed profile    in the
long \asca\ observation of MCG-06-30-15 \citep{Tanaka95},   were considered the first evidence   that
at least some line emission originates  in  the inner part of the accretion disk close  to the
central black hole. However,  the scenario emerging from  \xmm\ and \chandra\ observations of AGN
appears to be  more complex. Indeed, these observations   have shown  that only  a handful of
objects show the  presence of the relativistically broadened line, while the narrow Fe emission
line at 6.4 keV  is an  ubiquitous feature in many type I AGN (see \citealp{Bianchi04,Reeves04,
Yaqoob04}).  Furthermore, the broad component appears to be in general  weaker than  what was 
expected given the initial \asca\ results and in some case it may be absent (i.e. NGC~4151,
\citealp{Schurch03}).    These observations have also shown that the interpretation of  the Fe
profiles  can be strongly dependent  on the modeling of the  underlying continuum, which  can be
complicated by the presence of    complex absorption  (i.e. NGC~3783, \citealp{Reeves04};
NGC~3516,   \citealp{Turner05}) and reflection  components (see \citealp{Reeves07} and references
therein). Furthermore,  red- and  blue-shifted Fe absorption lines, associated with the  presence
of infalling or outflowing matter in the proximity of the black-hole, have been  detected in the
X-ray spectra of QSOs  and Seyfert galaxies (see \citealp{Cappi06} and references therein).  
These  absorption  and emission features, together with  the complexity  of modeling the underlying
continuum, makes the study of the Fe line profiles more complex and thus feasible only for  the
brightest objects. 

In this framework, MCG--5-23-16 represents one of the  best  and more robust examples of  a relativistically
broadened  Fe line. \sorg\ is a   nearby ($z=0.008486$) Seyfert 1.9 galaxy, with a  typical 2--10 keV flux of $\sim
8 \times 10^{-11}$ \flux\,    making it one of  the X-ray brightest Seyfert galaxies.    Previous X-ray observations
showed that  the  X-ray emission  of \sorg\ resembles the classical spectrum  of a Compton thin (i.e. $N_{\rm H} <
10 ^{24}$ cm $^{-2}$)  Seyfert 2, with a soft excess below 1 keV  and  a column density $N_{\rm H} \sim 10 ^{22}$
cm$^{-2}$ (\citealp{Dewangan03,Balestra}). Higher energy observations  (i.e. above 10 keV) performed with the {\it
Rossi X-Ray Timing Explorer} ({\it RXTE}; \citealp{Weaver98, Mattson04}) and {\it BeppoSAX}
(\citealp{Perola02,Risaliti02a}) detected  a Compton reflection component, that was interpreted as  reprocessed
emission from the distant molecular torus.  A strong broad Fe K$\alpha$ line was first detected with \asca\
 \citep{Weaver97,Weaver98} with  an equivalent width EW $\sim 200 $ eV, which could be modeled with a broad
relativistic line profile ($i\sim 50^\circ$, where $i$ is the disk inclination angle) plus a narrow  core at 6.4
keV (equivalent width EW $\sim 60$ eV). The presence of  both these   components has been subsequently confirmed
with   \chandra\    and  \xmm\ observations. In particular,   a narrow iron line component at $\sim$6.4 keV has been
clearly revealed with the \chandra\ observation; the intensity of this narrow core (EW $\sim$ 90 eV;
\citealp{Weaver01}) was found to be in  good agreement with the   \asca\ results. Meanwhile, two short \xmm\
observations, whose summed exposure time was $\sim $25 ksec,  confirmed the presence of a underlying broad
component  with an EW $\sim$ 100 eV   (\citealp{Dewangan03,Balestra}).  However, the  relatively short exposure time 
of the past \xmm\ observations  did not allow   the  above authors to put strong constraints on the  geometry of
the  emission region. Indeed    neither the inner disk  radius  nor the inclination angle could  be accurately 
derived using  only the \xmm\  data.    

\indent In this paper we  present an analysis  of the iron K line profile  and  variability   
from   simultaneous deep \xmm\   (130 ksec) and  \chandra\ (50 ksec)   observations of \sorg; 
the  analysis and results of the  simultaneous deep Suzaku ($\sim$100 ksec) are  described in
\citet{Reeves07}.  The  \xmm\ and \chandra\  observations and data reduction  are described in
section 2.  In section 3 we present the  modeling of the time-averaged  \xmm\  EPIC and RGS
spectra and the  results of the spectral fits of the simultaneous \chandra\  HETG spectra. In
section~4 we report the results obtained  with   time-resolved spectral analysis, aimed to assess
the possible  variability of the  iron K emission line during the long observation and  to
investigate the   relation (or lack of) between the Fe emission line intensity and the flux of the
underlying  X-ray continuum. In section 4, we  additionally discuss  the appearance    of a sporadic
absorption   feature  at  7.7 keV (rest-frame) that is indicative of a possibly variable, high
velocity (v $\sim 0.1\, c$) outflow.  The results are discussed and summarized in section~5.

\section{Observations and data reduction}
In December 2005, \sorg\ was observed simultaneously  with  many different X-ray observatories:
Suzaku, \xmm, \chandra\ and {\it RXTE};  in Table~1  we report the log of the different
observations. In this paper we  concentrate on  the \xmm\ and \chandra\ observations, while  the
Suzaku and {\it RXTE} observations  and results are  described in  detail by \citet{Reeves07}.  

\subsection {\xmm}
\sorg\  was observed with \xmm\ on 2005 December 8 for a total exposure time of about 130 ksec
(see Table~1). The pn, MOS1 and MOS2 cameras had     the medium filter applied; the MOS1 and
MOS2  were operating in Small Window mode, while the pn was in  Large Window mode. The \xmm\ data
have been processed and cleaned using the Science Analysis Software (SAS ver 6.5) and analyzed
using standard software packages (FTOOLS ver. 6.1, XSPEC ver. 11.3).  In order to define the 
threshold to filter for high-background time intervals,  we  extracted the 10--12 keV light
curves and   filtered out the data when the  light curve  is   2$\sigma$ above  its  mean.  This
screening yielded  net exposure times  (which also includes a dead-time correction)    of 96
ksec, 101 ksec and 103 ksec for the pn, MOS1, and MOS2 respectively.
 For the scientific
analysis  we  concentrated on the EPIC-pn data,  which have the highest signal-to-noise, and we
used the MOS1 and MOS2 data to check  for consistency. Taking into account  the brightness of the
source (the 2--10 keV  count rates  are 7.9 counts s$^{-1}$, 2.7 counts s$^{-1}$ and 2.8 counts
s$^{-1}$ for the pn, MOS1, MOS2, respectively), we ran the sas   task {\it epatplot} to check for possible 
pile-up and we found that both in the pn and MOS detectors the pile-up fraction is below  1\%.   
However, since we have good photon  statistics, when analyzing the  time averaged pn spectrum, 
we  decided to use only  the pattern zero data (which correspond to single events), which are 
better calibrated,   and we considered the pattern 0--4 (single and doubles) when we extract
spectra with lower exposure time for the time-variability study. 

The EPIC pn  source spectrum was extracted  using a circular region of $37''$ and  background data
were extracted using two circular regions with an  identical radius ($37''$) centered at $\sim 4'
$ from the source. EPIC MOS1 and MOS2 data were extracted using a source extraction region of
27$''$ radius and two background regions with identical size ($27''$)  selected on the nearby
CCDs.  Response matrices and ancillary response files at the source position  have been created
using the SAS tasks { \it arfgen} and {\it rmfgen}.  Background subtracted  data were then  binned
to have at least 50 counts  in each energy bin. \\The Reflection Grating Spectrometer (RGS;
\citealp{den Herder01})  data have been reduced using the  standard SAS task {\it rgsproc} and the
most recent calibration  files; the total exposure times are $\sim$ 97 ksec for  both RGS1 and
RGS2.  The RGS1 and RGS2   spectra were binned  at the resolution of the instrument ($\Delta
\lambda\sim 0.1\AA$).\\

\subsection{\chandra}
\chandra\  observed \sorg\   with the ACIS-S  with two relatively short exposures  for a total of
50 ks (see Table~1). For this study the \chandra\ observations were made with the High-Energy
Transmission Grating (HETG; \citealp{Markert94}) in the focal plane of the High Resolution
Mirror Assembly. The \chandra\ HETG consists of two grating assemblies, a High-Energy Grating
(HEG) and a Medium-Energy Grating (MEG); the  HEG affords the best spectral resolution in the
$\sim $6--7 keV Fe-K band currently available ($\sim$39 eV, or $1860 \rm \ km \ s^{-1}$ FWHM at
6.4 keV). The MEG spectral resolution is only half that of the HEG. The HEG also has higher
effective area in the Fe K band. The HEG and MEG energy bands are $\sim 0.9-10$~keV and  $\sim
0.4-8$~keV respectively, though  the effective area falls off rapidly with energy near both ends
of each bandpass. 

 The \chandra\ data were reprocessed with CIAO version 
3.2.1\footnote{http://cxc.harvard.edu/ciao} and  CALDB version 3.0.1. Spectral redistribution
matrices ({\tt rmf} files) were made with the  CIAO tool {\tt mkgrmf} for each arm ($-1$ and $+1$)
for the first order data of each of the gratings, HEG and MEG. Telescope effective area files were
made with the  CIAO script {\tt fullgarf} which drives the  CIAO tool {\tt mkgarf}. Again,
separate files were made for each arm for each grating for the first order. The effective areas
were corrected for the  time-dependent low-energy degradation of the ACIS CCDs using the option
available in the {\tt mkgarf} tool in the stated version of the  CIAO and  CALDB distribution.
Events were extracted from the $-1$ and $+1$ arms of the HEG and MEG using strips of width $\pm
3.6$~arcseconds in the cross-dispersion direction. Lightcurves and spectra were made from these
events and the spectral fitting described below was performed on first-order spectra combined from
the $-1$ and $+1$ orders (using response files combined with appropriate weighting), but keeping
the HEG and MEG spectra separate. The background was not subtracted as it is negligible in the energy
ranges of interest. Examination of the image of the entire detector and cross-dispersion profiles
confirmed that there were no nearby sources contaminating the data.

 In the following, unless
otherwise stated, fit parameters are quoted in the rest frame of the source  and errors are at the
90\% confidence level   for one interesting parameter ($\Delta\chi^2=2.71$). Abundances were set
to those of \citet{Wilms00}.

\section{Spectral analysis}
\subsection{The \xmm\ 0.3--10 keV continuum}
To characterize the X-ray continuum of \sorg\, we first fitted    the 2--10 keV pn data  with a  
redshifted power-law  model, modified by Galactic ($N_{\rm H}= 8\times 10^{20}$ cm$^{-2}$;
\citealp{Dickey1990})  and local  absorption.  For this initial fit, we  ignored the 5.0--7.5 keV
band, where the Fe K$\alpha$ emission is expected. This model yielded an acceptable fit
($\chi^2$/dof$=1066.3/987$) with  $\Gamma \sim 1.65$ and $N_{\rm H}\sim 1.3\times
10^{22}$ cm$^{-2}$. However,    the extrapolation of this model to the whole  0.5--10 keV band did
not provide a good fit ($\chi^2$/dof$= 6024.2/1794$).  Indeed, it left strong residuals in the
soft (E $<1$ keV)  band and, as expected, at the energy of  the Fe K$\alpha$ emission line.
Furthermore, it was not clear if the relatively flat  photon index was  intrinsic or indicative of   the presence of 
emission due to Compton reflection. To  illustrate this, in Fig.~1  we show the ratio of the pn
data to an absorbed power law model fitted over the 2-10 keV energy band (ignoring the 5.0--7.5
keV band), with the    photon index  fixed  to the  best fit value  $\Gamma=1.82$, derived with 
detailed modeling of  the \xmm\ and Suzaku broad band spectra (see Section 3.3). \\

\indent In order to  model the soft X-ray emission, we added  to this model a soft power-law
component absorbed only by the Galactic column density. The  photon index of this soft component
was found to be steep, $\Gamma= 3.13\pm 0.10$, and even at the CCD  resolution of the pn instrument,
the power-law model left line like residuals (see Fig..~2; black  data points). In particular 
an  emission line  is required  by the data  ($\Delta \chi^2=-37$) at $0.92\pm 0.02 $ keV with a
flux of $1.2\times 10^{-5}$ photons cm$^{-2}$ s$^{-1}$. This  emission line and the   steep
power-law continuum is confirmed by  the MOS1 and MOS2 data (see Fig.~2, red and green data
points)  and also by the simultaneous Suzaku observation \citep{Reeves07}. This  model gives a
2--10 keV observed  flux  of $\sim 8.2\times 10^{-11}$ \flux and an observed   luminosity of $\sim
1.3\times 10^{43}$ \lum.

This model is still too simple to describe the overall emission of
MCG--5-23-16, because it does  not account for the line emission seen in the RGS spectra (see
below).  However, it demonstrates that  scattering of the nuclear power law continuum is a
plausible explanation for the soft X-ray continuum spectrum.
  
\subsection{The RGS spectra: soft X-ray  spectrum dominated by emission lines}

In order to   investigate if   the apparent steep soft X-ray photon-index could be due to the presence of 
emission lines which are unresolved at the pn CCD  resolution, we analyzed the RGS data. Indeed, thanks to
the long exposure ($\sim 100$ ksec), the RGS1 and RGS2 spectra have   enough statistics to attempt a
spectral analysis (a total of  $\sim 17500$ net counts  between RGS1 and RGS2). The first inspection of
the RGS data reveals   the presence of several soft X-ray emission lines as well as the energy cut-off at
$\sim 1$ keV  due to the rest frame absorption. We  then rebinned the RGS spectra   in constant wavelength
bins at the  spectral resolution of the instrument ($\Delta \lambda\sim0.1\AA$) and  we used the {\it
C}-statistic  \citep{Cash79} available in
XSPEC\footnote{http://heasarc.gsfc.nasa.gov/docs/xanadu/xspec/manual/manual.html}   for the  spectral fit,
because  with this choice of binning, we  have  some bins with less than 20 counts.  

We first fitted the
RGS spectra  with the  baseline model obtained  with  the pn spectrum.  This model consists of  two
components:  a primary absorbed power law and   a scattered  soft power law component absorbed only by 
Galactic absorption; both   photon indices have been fixed to the value found for the primary power law
component ($\Gamma=1.82$, see Section~3.3). Overall, this model  provides a reasonable  description of the
RGS continuum, however, line-like residuals are present below 1 keV. We then kept this model, hereafter referred
as our   AGN baseline model and we tested two different additional  components for the emission
below 1 keV, which are:   ({\it a}) a multi-temperature  thermal emission model  with variable abundances
for different elements \citep{Mewe85} or  ({\it b})  several unresolved  emission lines.  The first model 
represents the emission expected from a collisionally ionized plasma; the latter reproduces the emission
due to   material photoionized by the AGN.  

When modeling with the multi temperature model, we found that the data could be fitted with two thermal
components with $k_{\rm B}T_1=0.44^{+0.19}_{-0.14} $ keV and $k_{\rm B}T_2=0.15^{+0.05}_{-0.06}$ keV ($\Delta
C=52$ with respect to the AGN baseline model,  for 9 additional parameters).  The flux and luminosity of these
thermal components are F$_{\rm (0.5-2) keV}\sim 9.7\times 10^{-14}$ \flux and L$_{\rm (0.5-2) keV}\sim 2.2 \times
10^{40}$ \lum, which are  consistent with  possible X-ray emission  from the galaxy. The data below 1 keV  still
required  a contribution  from the  scattered power law component of the AGN baseline model, which  has   F$_{\rm
(0.5-2) keV}\sim 2.1\times 10^{-13}$ \flux\.   Allowing the photon index of this soft power law  to vary, we 
found that it was no longer  unusually steep ($\Gamma\sim 2.1$). Although not well constrained, the abundances 
for N, O and Ne required  with this  model are found to be low,  $Z \sim 0.2 Z_\odot$ ($Z_N\sim 0.2\, Z_\odot$,  
$Z_O\sim 0.2\, Z_\odot$, $Z_{\rm Ne}\sim 0.4\, Z_\odot$). In particular  iron is found to be underabundant, with
only  an upper limit of  $ 0.2\,  Z_\odot$. This is due to the lack of Fe L shell emission lines and is   at odds
with the flux  measured    for Fe K$\alpha$ emission line (see Section 3.3 and Table~3). Furthermore this   value
is also  in contrast with the Fe abundance measured  with the  neutral  iron edge from the reflection component
in the simultaneous  Suzaku data ($Z_{Fe}=0.4\pm 0.1 \,Z_\odot $ at the 90\% confidence level;
\citealp{Reeves07}).   

We then fitted the  RGS spectra by adding to the AGN baseline   continuum model    several  unresolved emission lines, 
fixing both   the soft and the hard   power-law  photon indices   to the value  found for the AGN primary
power-law component  ($\Gamma=1.82$, see Section 3.3). To account for the excess of counts below 1 keV,  
five lines are required  ($\Delta C= 70$); their fluxes are listed in Table~2, together with their EWs
which range from $\sim$10 eV ({N\,\textsc{vii}})  to $\sim 46$ eV ({O\,\textsc{vii}}). The most likely
identifications are  2$\rightarrow$1 emission lines from  H and He-like O, Ne and N (see Fig.~3).  

We tentatively allowed the lines' width to vary and found  the  {O\,\textsc{vii}}  He$\alpha$, the {O\,\textsc{vii}}
radiative recombination continuum  (RRC), and  the {Ne\,\textsc{ix}} He$\alpha$ lines to be marginally resolved, while the
{N\,\textsc{vii}} Ly$\alpha$ and the  {O\,\textsc{viii}}  Ly$\alpha$ lines  are unresolved.  Although a  quantitative measure
is beyond the  statistics of the present data, from the width of the {O\,\textsc{vii}} RRC feature (E$> 0.739$ keV)  we 
derived  an upper limit of  $k_{\rm B}T$ $<$ 24 eV on   the recombining electron temperature.     It is worth noting that this low value
indicates  that the soft X-ray emission  originates in a photoionized  rather than in a collisionally ionized plasma
\citep{Liedahl96, Liedahl99}.   Taking into account the brightness of the  {O\,\textsc{viii}}   and  the {N\,\textsc{vii}}
Ly$\alpha$ line,   this photoionized material should also produce  RRC features from these ions. Neither  the 
{O\,\textsc{viii}}  (E$>$0.871 keV) or the {N\,\textsc{vii}} (E$>$0.667 keV) RRC are clearly detected in the  RGS spectra.  
However, the upper limit on the fluxes of both these features are fairly high ($5.9\times 10^{-6}$ph cm$^{-2}$ s$^{-1}$ and 
$3.8\times 10^{-6}$ph cm$^{-2}$ s$^{-1}$for {O\,\textsc{viii}}   and  {N\,\textsc{vii}} RRCs respectively). In particular  at
the energy of the  {O\,\textsc{viii}}  RRC feature the continuum of \sorg\  shows a steep rise due to the emergence
of the  absorbed power-law component, which also  complicates  a correct deblending of the {Ne \,\textsc{ix}} He$\alpha$ triplet (see
below).

The {O\,\textsc{vii}}  He$\alpha$ and the {Ne\,\textsc{ix}} He$\alpha$ lines are both triplets, which  with the present
statistics, can not be  resolved into their forbidden and resonance components. However it is  worth  noting that for both the
{O\,\textsc{vii}}  He$\alpha$  and the {Ne\,\textsc{ix}} He$\alpha$  lines the energy centroids are close to the energy of
the rest-frame forbidden line (see Table~2). This suggests a major contribution from  forbidden lines in each of  the
triplets with respect to the resonance lines. 

  In  the case of  {Ne\,\textsc{ix}} He$\alpha$, its   line energy is close to the   low energy   photoelectric cut-off present in the 
\sorg\ spectrum due to the local   absorption ($N_{\rm H }\sim 10^{22}$ cm$^{-2}$). Furthermore the  energy is also close to  the
{O\,\textsc{viii}} RRC ($>$0.871 keV) and to the Fe\,\textsc{xviii}-\,\textsc{xix} 3d-2p blend (0.853-0.926 keV). We then  included in
the model three more Gaussian lines to account for the   {O\,\textsc{viii}} RRC  feature   and  for the  decomposition of the 
{Ne\,\textsc{ix}} He$\alpha$ triplet\footnote{The energy  of the  forbidden (0.905 keV), the intercombination (0.915 keV)  and  the
resonance  (0.922 keV) components were fixed.}. The fit  did not statistically improve, and  we cannot derive any quantitative
information on the ratio between the  intensity of the forbidden and resonance line. In the case of  the {O\,\textsc{vii}} 
He$\alpha$ we measured  a width of $\sigma=4.6^{+8.9}_{-3.1}$ eV, which is probably due  to the presence of the intercombination  and 
resonance component. In order to confirm that    the detected emission is dominated by  the forbidden line, we then added two  lines
and  kept  the line energies frozen for the forbidden (0.561 keV), the intercombination (0.569 keV)  and  the resonance  (0.574 keV)
lines.  The measured ratio  between the  flux of the forbidden and resonance lines (for the latter we use the 90\% upper limit)  is
\simgt1.6; which again is  evidence of a strong contribution from  a  photoionized plasma \citep{Porquet00}. Furthermore with this
model we found that the  width of the forbidden line is now unresolved ($\sigma <4.7$ eV). Finally an   inspection of the values of
the  centroid  energies of  the detected lines (see   Table~2) shows  that  there are no strong shifts between the theoretical and
observed values. The measured shifts of these lines give a  value of \simlt1 eV (which  for example for the {O\,\textsc{viii}}
Ly$\alpha$    corresponds to a velocity \simlt 500 km s$^{-1}$). 

In order to test whether these lines can be  explained with  emission from  optically thin gas photoionized by the AGN,  we replaced
the  unresolved emission lines with a grid of photoionized emission models generated by  {\sc xstar} \citep{Bautista01}, which assumes
a $\Gamma\sim 2$  illuminating continuum and a turbulence velocity of $\sigma_{\rm {v}}=100$ km s$^{-1}$. We found that the RGS data
are well explained  with this model with an ionization parameter log$\xi = 1.29_{-0.13}^{+0.17}$.  We  then  let the photon index of
the scattered power-law vary and we found that, although it is not well constrained,  the value is  now similar
($\Gamma=1.97^{+0.18}_{-0.53}$) to the primary AGN power law emission, in agreement with the scattering hypothesis.  Though  even with
this model  the abundances cannot be constrained,    iron ($Z_{\rm {Fe}}=1.1^{+1.2}_{-0.8}Z_{\odot}$) is not underabundant with respect
to the value  obtained  with the detailed fitting of the Fe  K$\alpha$ line and the Compton reflection hump detected with the Suzaku
observation \citep{Reeves07}.

 As a final test, we   applied this best fit model to
the pn, MOS1 and MOS2 data. We kept the abundances  fixed to the values measured with the RGS spectra  and we let only the
normalization    and the photon index  vary. This model is now a good description of the soft spectrum and no strong
residuals are  present. Finally with this model we found that the scattered component has a  
$\Gamma=2.3\pm0.3$ and F$_{\rm (0.5-2) keV}\sim 4\times 10^{-13}$ \flux. This corresponds to 0.5\%  of the
un-absorbed  flux of the primary AGN component and  the emission due to the photoionized gas (L$_{\rm (0.5-2) keV}\sim 
10^{40}$ erg s$^{-1}$) is 0.1\% of the  AGN emission.\\ 

To conclude, although from a statistical point of view either the
multi-temperature  thermal emission model   or the photoionized plasma model ({\sc xstar})  give similar results, 
different  diagnostics   suggest  that, as  already seen in  other Seyfert2 galaxies \citep{Bianchi06,Guainazzi07,Iwasawa03} the soft X-ray emission is likely due to photoionized plasma which could be  associated  with the Narrow Line
Region.  The properties of the soft X-ray emission of \sorg\ that favor this interpretation are: the lack of a strong
Fe-L shell emission \citep{Kall1996}, the detection of a narrow RRC feature from  {O\,\textsc{vii}} and the  stronger contribution from the
forbidden line in the  {O\,\textsc{vii}}  He$\alpha$ triplet.\\

In order to assess the extent of the soft flux, we  used  the \chandra\ observations   and created an image  combining the 0.3--1.0 keV photons
from the zero-order of both  observations.  We created a point-source function (PSF) using the
MARX\footnote{http://space.mit.edu/CXC/MARX/} \chandra\ simulator.  We then fit a model  to the \chandra\ image consisting of a constant
component, to account for  the background, and two Gaussian components with the centroid positions  tied together, to account for
both nuclear and extranuclear emission. The  model was convolved with the PSF and then compared to the data using  the Cash
statistic.   Initially the image was fit with $\sigma_x  = \sigma_y$, i.e., circular Gaussian models.  This gave  $\sigma =
0.39''$$^{+0.23}_{-0.15}$ for the nuclear component and $\sigma = 1.57''$$^{+0.78}_{-0.31}$ for the extranuclear component.  The
count rates for the  two components were $1.4 \times 10^{-3}$ counts s$^{-1}$ and $1.2 \times  10^{-3}$ counts s$^{-1}$.  Note that
some fraction of these  extents is likely due to residual error in the aspect solution,  which would effectively increase the PSF. 
Unfortunately there are  no other on-axis point sources bright enough in the field to check  this.  Allowing the extra-nuclear
component to  be elliptical did not improve the fit significantly, however it reduced $C$
by 7.2 for 2 additional  parameters (the additional $\sigma$ parameter and the rotation angle), which is significant at the 2$\sigma$
level.  It is likely,  however, that this asymmetry may be due to aspect errors.  As a  consistency check we fitted the zero-order
image from the 2000  \chandra\ observation with the same model.  In this case allowing  for ellipticity in either component did not
improve the fit significantly.   In this fit the best-fitting parameters were $\sigma = 0.56''$$^{+0.09}_{-0.12}$  for the nuclear
component and $\sigma = 2.2''$$^{+0.5}_{-0.6}$ for the extra-nuclear component, which is consistent within the errors. This is
indicative that  around half of the soft X-ray emission is due to the central point like source and half to an  extended component.
Assuming the current cosmology (H$_0=71$ km s$^{-1}$ Mpc$^{-1}$, $\Omega_{\Lambda}$=0.73, and $\Omega_m$=0.27)  the soft X-ray
emission of this latter component  originates within  $\sim 0.7$ kpc; this value is  in
agreement with a possible association with the NLR; indeed this extension  is comparable to the extension of the
$[{\rm {O}\,\textsc{iii}}]\lambda 5007$  derived with  HST data   \citep{Ferruit00}.\\

\subsection{The iron  K band}
 
 We then considered the hard X-ray emission of MCG--5-23-16, using  the dual power law continuum as described above and  only the 
Gaussian emission line at $\sim 0.9$ keV.  In Fig.~4  we show the residuals left by the absorbed power law model (with $\Gamma=1.65$)
at the energy of Fe K band. These residuals  clearly reveal the presence of a   strong narrow core at the expected energy of the Fe
K$\alpha$ (6.4 keV)   and broad wings,   which   extend from $\sim 5.7$ keV to $\sim 7$ keV. The pn data also show    a narrow emission
line  at $\sim 7$ keV,  due to  Fe K$\beta$  and a drop at 7.1 keV probably due to a reflection edge. The presence of this latter
component  was already  suggested   with the previous short \xmm\ observations \citep{Dewangan03}, however  the   short  exposure of
these observations together with the lack of any  simultaneous  observation above 10 keV did not allow the authors to  put strong
constraints on this feature.  We adopted   the best-fit model obtained  by \citet{Reeves07} from the simultaneous Suzaku observation
for the underlying continuum  in order to  derive  the Fe K$\alpha$ emission line properties. Indeed Suzaku's  broad band  energy range
(0.4--100 keV) allowed these authors to   measure the amount of Compton reflection   and thus better   constrain the    continuum in
the  2--10 keV energy band. This  model is composed of: a primary absorbed power-law component with an exponential cut-off at high
energies ($>200$ keV) and  a component   due to reflection from  neutral  material (the  PEXRAV model in XSPEC, \citealp{pexrav}). The 
parameters of this  reflection component are: the  reflection fraction,  which is defined by the subtending solid angle of the
reflector $R=\Omega/2\pi=1.1$; an inclination angle $i=45^\circ$ and abundance $Z=0.4Z_\odot$.  When fitting this model to the pn data
we  kept    the values of  $Z$, R and    the  cut-off energy fixed, since they cannot be determined   using the lower energy band pass
of \xmm. After including  the reflection component, the residuals  no longer  show the deep edge at 7.1 keV, which   is  well  modeled
with the  reflection component (see Fig.~5; upper panel). With this model we found that the primary  power-law component has a photon
index $\Gamma=1.82\pm 0.01$,  absorbed by  a neutral column density of $N_{{\rm H}}=1.49\pm 0.01 \times 10^{22}$ cm$^{-2}$.\\ To model
the Fe line we  first added   narrow Gaussian lines at the energies of Fe K$\alpha$  and  Fe K$\beta$. For this latter line  we kept
the energy fixed at 7.06 keV and  tied its flux to be 12\% of the Fe K$\alpha$ flux. This model  clearly leaves  an excess  of counts
at the energy of the 6.4 keV Fe K$\alpha$ line (see Fig.~5; middle panel),  which can be accounted for by including a  broad Gaussian
line  or  a relativistic disk-line component (see section 3.3.2).\\

\subsubsection{\chandra\ Observation of the Narrow Core.}

In order to measure  the  parameters (i.e. strength and profile)  of the broad component we first
derived the width and flux of the narrow core using the simultaneous \chandra\ observations. We   
combined the  \chandra\ $\pm 1$ MEG and HEG 1st order spectra of both   observations. The  
combined  spectra were rebinned at the maximum spectral resolution of the instruments ($\Delta
\lambda=0.012\,\AA$   and $0.023\,\AA$   for HEG and MEG respectively) and the   spectral fit was
minimized with the {\it C}-statistic  \citep{Cash79}.\\  
  We then adopted the Suzaku best fit model 
for the underlying  2--8 keV continuum. Thanks to the high  resolution of the MEG and HEG 
instruments, the \chandra\ HETG  residuals clearly reveal   the presence of a narrow core at 6.4
keV (see Fig.~6), which is best modeled with a Gaussian line at $6.40^{+0.02}_{-0.01}$ keV and 
EW$\sim 80$ eV ($\Delta C= 43$). With this model we measured a    width of $\sigma=32^{+19}_{-16}$
eV, which corresponds to a velocity width of $\sigma_{\rm{v}}\sim 1400 $ km s$^{-1}$  and is thus
indicative    of  a possible origin from the molecular torus. \\ Taking into account that the   
measured width of the  Fe K$\alpha$  narrow core could be due to the presence of the broad
component, we added a  second  Gaussian line.  Although  the fit  did not statistically improve, 
we found an  EW of $\sim 60$ eV (flux $\sim 5.6 \times 10^{-5}$ photons cm$^{-2}$ s$^{-1}$;
$\sigma\sim 0.4$ keV) and $\sim 70$ eV (flux $\sim 5.8 \times 10^{-5}$ photons cm$^{-2}$ s$^{-1}$) for
the  broad  and narrow components  respectively,  which  are in good agreement with the values
measured with the \xmm\  (see Table 3) and Suzaku spectra  \citep{Reeves07}.  With this model
the narrow core is no  longer  resolved and we can place only an  upper limit  on the width of 
$\sigma < 43$ eV, which corresponds to a FWHM $< 5000$ km s$^{-1}$ (at the   90\% confidence level) in 
good agreement with the  upper limit measured with a previous \chandra\ observation of \sorg\
(\citealp{Balestra}). \\

\subsubsection{The broad Fe line}

We then adopted the \chandra\ upper limit on  the width of the narrow core for the \xmm\ fits.   The  broad component    was
first modeled adding  a   second  Gaussian line. We also added a Compton shoulder at 6.3 keV, with its normalization set to
20\%  of the Fe  K$\alpha$ flux \citep{Matt02};  the fit improved with a  $\Delta \chi^2 = 44$ for 3 additional parameters
($\chi^2=1932$ for 1785 dof). The  broad Gaussian component (E = $6.22^{+0.11}_{-0.16}$ keV) has  an EW of $66^{+19}_{-18}$ eV
and a width of $\sigma=0.42^{+0.14}_{-0.10}$  keV;  which corresponds to a velocity of $\sigma_v\sim 20000$ km s$^{-1}$ (FWHM$\sim
40000$ km s$^{-1}$).
 
 We then tested a relativistic diskline model   (DISKLINE  in XSPEC; \citealp{Fabian1989}); this code models a line  profile
from an accretion disk around a Schwarzschild black hole.  The main parameters of this model are the inner and outer radii of
the emitting region on the disk, and its inclination. The  disk radial emissivity is assumed to be a  power law, in the form of
$r^{-q}$. For the fit we  fixed the  outer radius to be 400R$_{\rm g}$  (with $R_{\rm g}=GM/c^2$) and the emissivity   $q=3$.
Finally we assumed the   line to be from neutral Fe K$\alpha$. From a statistical point of view this fit  gives a similar
result to the model with a broad Gaussian line  ($\chi^2=1928$ for 1785 dof), however  the  high velocity inferred from the
width of the Gaussian line is indicative that the line must be produced close to the  central black hole; i.e.    within
100R$_{\rm g}$  and thus inside the Broad Line Region.  With this model we found that the inner radius is $R_{\rm
{in}}=48^{+62}_{-20}$R$_{\rm g}$, and the inclination angle is $i=41^{+29}_{-12}$$^\circ$; while  the EWs of the   broad  and
narrow components are  EW$_{\rm {Diskline}}=53^{+14}_{-13}$ eV and EW$_{\rm N}=64\pm 6$ eV respectively. If the constraint on
the  disk emissivity is relaxed and a flatter emissivity is assumed ($q=2$), then a disk inner radius of $6R_{\rm g}$ is 
allowed by the present data. The ratio between the data and this best fit model is shown in Fig.~5 (lower panel), an
absorption  line near 8 keV is the only  residual. Adding an absorption line  to our best fit model  improved the fit
($\Delta \chi^2=20$ for 2 additional parameters, $|EW|\sim 30$ eV, E$\sim$ 7.9 keV).  In Fig.~7 the  resulting EPIC-pn spectrum
and the  best model components are shown. The parameters  derived for the diskline did not change significantly; the  main
difference is a slight reduction in its EW, which became now $46^{+14}_{-13}$ eV. Taking into account that this absorption feature
is indicative of  the presence of an ionized absorber (see section 4.1 and 5.3), we  tested if the presence of a more complex
absorber could mimic the profile of the  detected broad   component. We found that    the inclusion of a two   layers of
absorption, characterized by a  high (log$\xi=3.7$)  and low (log$\xi=2$)  ionization level, did not   impact  the
detection  of the broad component and its  parameters.\\

\section{Variability of the Iron line and continuum} 

During the \xmm\ observation, the 2--10 keV flux of \sorg\ varied from $\sim 7\times 10^{-11}$\flux  to $\sim
9\times 10^{-11}$ erg~cm$^{-2}$~s$^{-1}$. In order to investigate the possible variability of the line properties
and    continuum shape,  we first tested if there was any clear  difference between  the spectra extracted    when
the source was in relatively  higher and lower flux states. We extracted spectra using  2--10 keV threshold of $<4.6 $ counts s$^{-1}$ (F$_{\rm{(2-10\; keV)}}\simlt 7.4\times 10^{-11}$ \flux) and $>5.5$ countss$^{-1}$  (F$_{\rm{(2-10\; keV)}} \simgt 8.8\times
10^{-11}$ \flux).  We then fitted  both spectra with the previous best-fit model, replacing the diskline with a
broad Gaussian. We found  no evidence of variability of either the broad or  the narrow component; indeed their 
normalizations are consistent within the errors (see Table 3). Furthermore, the width of the broad line is  
constant.  In order   to confirm that the line is not strongly variable, we checked the difference spectrum,
obtained by subtracting the low from the high state data. The  difference spectrum can  be    modeled
with an absorbed power law with photon index $\Gamma=1.80\pm 0.09$; no strong residuals are left in either  the soft
band or  in the iron K band.

As a second test  to assess the possible  variability of the Fe emission complex, we divided the \xmm\ observation
into 5   intervals with a duration of  20 ksec each; for the fit we tied all model parameters  except the
normalization of the primary power-law component. This model gives a statistically acceptable  fit for all the  5
spectra.  Fig.~8  shows the  5.5--8.5 keV residuals  to this model for all  five intervals;  no strong deviations
from the model are required  at the energy of the iron K$\alpha$ line. We then allowed the normalizations of the
narrow and broad components free to vary.  Fig.~9 (panel a and panel b)  shows the fluxes  of  both the  
narrow and   broad components. There is no evidence of variability during the present observation;  furthermore, the
fluxes of  both components are consistent within the errors  with the values measured  in the previous \xmm\ and
\chandra\ observations \citep{Dewangan03,Balestra}. This lack of variability can be explained by taking into account
that \sorg\ is not highly variable on either  short and relatively long time-scales; indeed the source has remained at a
similar flux level  (7--9 $\times 10^{-11}$ \flux) for the last 10 years.

 The most striking  feature that
appears to be  variable during the \xmm\ observation is a possible absorption line at $\sim 7.7$ keV (E $\sim 7.66 $
keV, observer frame). This feature is present in  the average EPIC pn spectrum but as shown in Fig.~8, it  is
strongest  in the third  spectrum (40 ksec after the beginning of the \xmm\ observation). To illustrate this  in
Fig.~9 (panel c) we compare the  intensity  of this absorption feature during the five intervals. For this purpose
we modeled  the absorption with an inverted Gaussian line and fixed the energy to the best-fit value found with the
spectral analysis of  the third interval (E $\sim$ 7.7 keV, $|I|\sim 3.2 \times 10^{-5}$ photons cm$^{-2}$ s$^{-1}$,
EW $=52\pm15$ eV). The line is clearly variable and it  appears to be strongest  during the  interval with the
slightly higher 2-10 keV flux (see Fig. 8, panel d), while  it is  barely detected in the other four spectra.

\subsection{A variable absorption feature at 7.7 keV}
\subsubsection{Epic-pn background and  calibration checks.}

Before attempting any further modeling, we performed several tests to exclude that  the 7.7 keV    absorption
feature is  due to inappropriate background, binning or pattern selection.\\ 
The EPIC-pn background  near this
energy range  presents two instrumental lines due to Cu (8.05 keV) and Ni (7.48 keV) K$\alpha$  emission lines and 
an inadequate background selection  could  in principle  cause spurious    absorption features. However several
arguments exclude this possibility. First of all the  net count rate ($\sim 2.56$ counts s$^{-1}$) of \sorg\  in the 5-10
keV  band  is $\sim 300$ times  greater than the background  ($\sim 8\times 10^{-3}$ counts s$^{-1}$). Second,  the sporadic
nature of the  feature is indicative that the feature can not be an artifact of the background or calibration of
the EPIC-pn. Finally   there was no background flaring activity during this time interval. We  conclude that the
feature is not due to instrumental or  external background. 

In order to  exclude the possibility that the 7.7
keV feature is due to a  binning effect we rebinned the pn data of the  third interval with a constant energy binning
of 80 eV. This choice  corresponds to about half of the energy resolution  of the EPIC-pn camera in this energy range
(FWHM $\sim 150$ eV at 6.4 keV; see the XMM-Newton Users' Handbook, Ehle et al.
2006\footnote{http://xmm.vilspa.esa.es/external/xmm\_user\_support/documentation/uhb/XMM\_UHB.html}). As shown in 
Fig.~10, the residuals left by the time averaged best fit model confirm the presence of the feature,   thus
excluding  the possibility that it is an artifact  of the  choice of the  binning. \\ 

To exclude a pattern
selection effect we  then compared the pn spectra  extracted with the pattern 0-4 and pattern 0 selection criteria.
Though the latter has   30\% fewer counts we found no significant difference  in   the absorption line parameters
($\Delta \chi^2=26$ for 2 dof; E$\sim 7.7 $ keV, EW$\sim 60$ eV).
Finally the presence of an  absorption  feature is confirmed by the MOS1 and MOS2 spectra extracted  in the same
time interval.  Although,  due to the lower photon statistic,  the significance of the absorption line is lower in
these spectra,   both  the  flux and the energy of the feature are consistent ($|I|= 2.2\pm 1.8 \times
10^{-5}$ photons cm$^{-2}$s$^{-1}$ E =$7.4\pm0.2$ keV $\Delta\chi^2=6$; see Table~4) with the  values found with the pn
data\footnote{Unfortunately the \chandra\ observations do not overlap with this segment of the \xmm\ spectrum.
However two  possible weak absorption features appear to be present at the rest frame energy of about 7.3 keV and
7.4 keV (see Fig.~6) suggesting  possible variability of the absorber, although the statistical significance of
these features is low. }.  Furthermore the presence of the absorption feature is  confirmed
by the simultaneous Suzaku observation, indeed  a weak absorption feature is present in  the  time averaged spectrum
(see panel c of Fig.~6 in Reeves et al. 2007). The absorption  line is weaker than in the \xmm\ observation, which could
be explained if we take into account  the apparent sporadic nature of the feature,  with a dilution   effect due to the
longer duration time ($\sim $220 ksec) of the  Suzaku observation in the  Earth orbit.  Although the absorption line
is not well constrained in the time-averaged Suzaku spectrum, the energy of the line at $7.8\pm0.1$\,keV is coincident
with the \xmm\ data, while the flux of the line is weaker 
$|I|=1.8\pm0.9\times10^{-5}$\,photons\,cm$^{-2}$\,s$^{-1}$ (see Table~4).

\subsubsection{Modeling the absorption feature of the third segment.}
We first fitted  the absorption feature adding a Gaussian shaped absorption line keeping its width 
fixed to $\sigma=0.1 $ keV. The addition of this line improved the  fit  with a
$\Delta\chi^{2}=33$ for 2 additional parameters ($\chi^2/dof$=1451/1505). The line energy is $7.72\pm 0.06$
keV with a $|$EW$|$ of $52\pm15$ eV ($|I|= 3.2\pm0.9 \times 10^{-5}$ photons cm$^{-2}$s$^{-1}$).  Leaving the
width of the line free did not improve the fit  significantly ($\Delta\chi^{2}=5$ for one additional
parameter). With this fit we  find  $\sigma=0.2\pm 0.1$ keV,  EW=$78\pm29$ eV and an energy
consistent with the previous best fit (E=$7.71\pm 0.08$ keV).  In Fig.~11 we show the confidence contour
plot  of the line parameters (rest frame energy and intensity) with  the line width left free to vary.   
We also attempted to fit the  absorption feature  replacing the  Gaussian line with an edge due to K-shell
absorption from  partially ionized iron. This model gives a best-fit energy  of 7.33$^{+0.12}_{-0.22}$ keV and an optical
depth of $\tau=0.09\pm 0.03$. The fit   is statistically acceptable; however, it is   worse than the  Gaussian
absorption  model ($\chi^2/dof$=1465/1505, which correspond to a  $\Delta\chi^2=14$ worse compared to the 
absorption line)  and more importantly, it is unsuccessful  at  modeling the residuals at 7.7 keV.\\ 

\subsubsection{The significance of the detection of the absorption line}
By applying the standard two-parameter  $F$-test to  the drop in  $\chi^{2}$ of 33   for the addition of an inverted
Gaussian   at 7.71 keV,   we found  a null hypothesis  probability  for adding this extra component   of $\sim4\times
10^{-8}$. However, as discussed  by  \citet{Protassov02}, the $F$-test applied  in this  way  could overestimate the
true significance of the detected absorption line. In particular the $F$-test does not take into account the  number
of time bins in which  the line is searched  as well as the range  of energy where the line might be expected (see
\citealp{Porquet04}).

  To asses the significance of the detection  we then performed Monte Carlo simulations as
described  in \citet{Porquet04} and in  \citet{Markowitz06}, for a similar case of a blue-shifted Fe
K$\alpha$ absorption line detected in the \xmm\ observation of IC~4329a. We assumed  as our null hypothesis model 
the best fit model with no absorption feature, and we simulated  1000 spectra  with the photon statistics expected
for a 20 ksec exposure.   Each simulated spectrum was then  fitted  with the null hypothesis model  to obtain   a
$\chi^2$  value and we systematically searched for an absorption line over the 4--9 keV energy range stepping the energy centroid
of  the Gaussian in increments of 0.1 keV, refitting at each step.  We then  obtained for each
simulated spectrum a minimum  $\chi ^{2}$ and  created a distribution of 1000 simulated values of the $\Delta
\chi^{2}$ (compared to the null hypothesis model), which was used to construct a cumulative frequency distribution
of the $\Delta \chi^{2}$ expected for a blind line search in the 4--9 keV range.  Not a single fake spectrum had
a     $|\Delta \chi^{2}| \ge  33$, thus  the inferred probability that the null hypothesis model was correct is
$<0.1$\%. Taking into account the number of   intervals (5) into which the observation had been split,   we  derive that the line detection
is significant at $> 99.5$\%.

Finally, we performed   a  similar Monte Carlo simulation  to test  the  statistical significance of the
absorption feature in the  MOS spectra. For simplicity we    ran the simulation on MOS1 only and we found that  in this case the significance 
is only $\sim 61\%$,  mainly due to the lower  S/N of the MOS spectra at this energy. However the  fact that the absorption feature is
detected by the pn, both MOS cameras and Suzaku suggests that the feature is likely
real and not an artefact.\\ 

\subsubsection{The ionized absorber model.}
As already discussed from  X-ray spectroscopic observation of several other Seyfert
galaxies (i.e. MCG--6-30-15,\citealp{Young05}; NGC~3783, \citealp{Reeves04};
Mrk~509, \citealp{Dadina05}; Mrk~335 \citealp{Longinotti07}; E1821+643,
\citealp{Yaqoob05};  IC~4329a, \citealp{Markowitz06}) and QSOs  (i.e. PG~1211+143, 
\citealp{Pounds03};  PDS~456, \citealp{Reeves03}; APM 08279+5255, \citealp{Chartas02};
PG~1115+080, \citealp{Chartas03}) a likely candidate for the 7.7 keV  feature is
blue-shifted  K-shell  absorption  due to  He- or H-like iron.  In particular if we
assume that the line is due to H-like iron ({Fe\,\textsc {xxvi}} Ly$\alpha$ at 6.97
keV) the observed blue-shift suggests that the absorber is outflowing with a
velocity of the order of 0.1$c$.   

In order to obtain a more physical representation of the  absorber, we replaced the Gaussian line with a model  comprised of a
grid of photoionized absorbers generated by the {\sc xstar} photoionization code (\citealp{Bautista01}). For the absorber we
assumed  a one zone  photoionization model with half solar abundances  and a  turbulence velocity of 1000 km s$^{-1}$. The free
parameters of this model are: the column density ($N_{\rm H}$), the outflowing velocity of the absorber ($v_{\rm out}$) and the
ionization parameter ($\xi= L/nr^2$; where $L$  is the  ionizing luminosity, $n$  is the electron density, and $r$ the 
absorber distance).  To reproduce the absorption feature  a column density  of $ \sim 8\times 10^{22}$ cm$^{-2}$ and an
ionization  state characterized by log$\xi=3.7^{+0.2}_{-0.3}$ erg cm s$^{-1}$   are required with an outflow velocity of
$0.09\pm 0.01\,c$ ($\sim 30000$ km s$^{-1}$).  A plot of this best-fit model is shown in Fig.~12 (model A). The column density
is not well constrained and we can place only a lower limit of $>2 \times 10^{22}$ cm$^{-2}$. At this ionization level, the Fe
K-shell absorption is indeed mainly due to {Fe\,\textsc {xxvi}} and is consistent with absorption from highly ionized iron
outflowing at $\sim 0.1\;c$ with respect to systemic.\\

A lower velocity outflow could in principle be obtained assuming that the feature is due to a {Fe \,\textsc {xxiv}}
1s-3p line at 7.78 keV. This is illustrated in Figure 11 (model B), which illustrates an {\sc xstar} model with a column
density of $N_{\rm H}=10^{23}$\,cm$^{-2}$, an ionization parameter of $log\xi=3.0$ erg cm s$^{-1}$ and no velocity shift. The absorption line
at 7.8 keV indeed corresponds to the above {Fe\,\textsc{xxiv}} $1\rightarrow3$ transition. However, in this ionization
regime ($log\xi\sim 2.5-3.0$), we would expect to detect   absorption trough  at  6.5-6.7 keV due to a blend of the $1\rightarrow2$
transitions of {Fe \,\textsc{xviii-xxv}}; however, this is not observed in the MCG\,-5-23-16 spectrum. In
particular, at this lower ionization we would expect to  see the strong absorption due to the {Fe \,\textsc{xxiv}} 1s-2p
absorption line at 6.67 keV, and  for  $log\xi=3.0$ erg cm s$^{-1}$ (as illustrated in the figure) we would also expect
to see  an even deeper absorption feature at 6.7 keV due to   the {Fe \,\textsc{xxv}} resonant absorption.  Furthermore in this ionization regime ($log\xi\sim 2.5-3.0$), several strong absorption
lines from iron L-shell ($2\rightarrow3$) transitions as well as He
and H-like Si/S K-shell lines are expected near 1--2\,keV, which are
not observed in either the \xmm\, Suzaku, or \chandra\ HETG
spectra. Therefore this lower velocity solution appears to be ruled
out.

Finally it is possible to have a low ionization K$\beta$ absorption
line without strong K$\alpha$ absorption from the same species. This scenario is shown in Figure 11 (model
C) for an absorber with a column density of $N_{\rm
H}=10^{23}$\,cm$^{-2}$ and an ionization parameter of $log\xi=1.5$ erg cm
s$^{-1}$.  Indeed, at  this ionization state the dominant species is iron less ionized then {Fe \,\textsc{xvii}};   
the   L-shell is  full and  cannot produce the   $1s-2p$ absorption between 6.4--7.0\,keV. However, there is an
absorption line from 7.1--7.2\,keV due to $1\rightarrow3$ transitions
from Fe less ionized then {Fe \,\textsc{xvii}}. In this scenario a blueshift of
$\sim0.08c$ would still  be required to model the absorption line at
7.7\,keV in the spectrum. Furthermore the column density 
required to model the EW of the K$\beta$ absorption feature is
$N_{\rm H} >> 10^{23}$\,cm$^{-2}$, which would introduce too much
continuum (bound--free) absorption below 6 keV, inconsistent with the
pn data.\\  Therefore we conclude that the fast (0.1$c$) high
ionization outflow is the most likely model to account for the
absorption feature at 7.7\,keV. Moreover, when we compare the  2--10
keV continuum and the neutral $N_{\rm H}$ measured  during the third
interval with the other intervals and with the average spectrum, we
do not find any statistically significant difference ($\Delta N_H <
10^{21}$ cm$^{-2}$), which rules out the possible presence of 
variable neutral and/or low ionization absorber. \\

\section{Discussion and Conclusions}
We have presented the results  from \xmm\ and \chandra\ observations of \sorg, which
are part of a simultaneous   campaign  conducted in December 2005 also comprising 
Suzaku  and {\it RXTE} observations.   The 0.5--10 keV  continuum   of \sorg\   resembles at first
order the canonical  X-ray emission expected from a Compton thin Seyfert 2 galaxy:
an  absorbed  ($N_{\rm H}=1.5\times 10^{22}$ cm$^{-2}$) power law component ($\Gamma=
1.82$), which emerges at   energy  $\simgt$1 keV, and   a steep soft  excess, which
is well fitted by a power law    component plus several emission lines from O, Ne,
and N. The  \xmm\ observation of \sorg\  confirms the presence of the  Fe K$\alpha$
emission complex, which is well described by a narrow  Fe K$\alpha$ emission line
superimposed on a   relativistically broadened  component.   The  simultaneous
Suzaku observation   provided us with an accurate  description of the underlying
continuum which allowed us  to perform detailed modeling  of the Fe K emission line
complex.  Finally due to the sufficiently long duration,  the \xmm\ RGS spectra 
have  enough photon  statistics to investigate the origin of the soft X-ray
emission.\\

\subsection{The origin of the soft X-ray emission}
The analysis of the EPIC-pn and MOS spectra  of \sorg\ revealed the presence, below
1 keV,  of a soft excess with respect to the  primary nuclear emission. This soft
excess  can be well fitted  by adding an un-absorbed power law component to the primary
AGN emission, the photon index of this power law  is  found to be steeper
($\Gamma\sim 3.1$) than the primary AGN component ($\Gamma\sim 1.8$), and even at
the EPIC CCD resolution  an emission line is detected around 0.9 keV. 

 A soft
excess  below 1 keV is not unusual  in obscured Seyfert galaxies like \sorg\ (see
\citealp{Bianchi06} and references therein) and  it   has been already suggested
that it could be due to a superimposition of     scattered emission into the line of
sight by  ionized gas plus several emission lines from highly  ionized  (He and H
like)  elements  (i.e Mrk~3: \citealp{Sako00,  Pounds05, Bianchi05};  Circinus: 
\citealp{Sambruna01}; NGC~1068: \citealp{Kinkhabwala02, Ogle03, Brinkman02};
NGC~4507: \citealp{Matt04}). Key diagnostics to
understand the origin of this X-ray emission when  a high resolution spectrum   is
available are: the  detection of  RRC transitions,  the detection of  enhanced
K-shell  emission lines from H-like and He-like   ions, the detection  of Fe L-shell
emission  and the   ratio between  the forbidden  and the recombination transition
in the He-like triplets.  

This long \xmm\ observation has provided for the first time   RGS spectra for \sorg\  with  sufficient  photon statistics to
perform  a detailed  modeling of the soft X-ray emission, allowing for the first time in this object the detection of
the {O\,\textsc{vii}} RRC. The  width  of this emission line  indicates  that the recombining electron temperature   is
a few eV ($k_{\rm B}T <24$eV). This is suggestive that the emitting plasma is photoionized rather then collisionally ionized
\citep{Liedahl96}. This is also indicative that the soft X-ray emission is probably   dominated by scattering of the
primary AGN emission rather than due to  emission   from  hot gas  in the host galaxy , e.g. from  starburst
activity.  Our analysis of the RGS spectra of \sorg\ confirms the detection of the {O\,\textsc{vii}} and
{O\,\textsc{viii}}  lines previously reported by \citet{Guainazzi07} and the line fluxes are in agreement  with the
measurement obtained   with the analysis of these previous short \xmm\  observations of \sorg.  We  cannot exclude on a
statistical ground the possible presence of emission due to a collisionally ionized plasma. Indeed the  spectra can be
equally modeled by replacing these emission lines with   a multi temperature thermal model; which  represents the emission 
due to a collisionally  ionized plasma. However the  ``AGN"  model (scattered power law component plus several 
photoionized emission lines)  is preferred because of the low ion abundance obtained ($Z<0.2 Z_\odot$) in the  thermal
model, which reflects the lack of  a strong Fe L-shell emission  with respect to the fluxes of the Oxygen lines.  \\ We
therefore conclude that  the most likely the origin of the soft X-ray emission is due to a plasma photoionized by the
AGN. This plasma must be  located  outside the Compton-thin absorber and, as already suggested for other Seyfert 2 
galaxies, it could be coincident with the NLR.\\

\subsection{The Fe K$\alpha$ emission complex}
This deep \xmm\ observation of \sorg\ confirms the presence of  broad  and  narrow  iron K$\alpha$ emission
lines, which were  reported since the first \asca\ observation \citep{Weaver97}. The \chandra HETG  spectrum
clearly reveals a   narrow line at $E_N=6.40\pm0.02$ keV with a FWHM $< 5000$ km s$^{-1}$ and  a flux of
$5.6\pm0.7\times 10^{-5}$ photons cm$^{-2}$ s$^{-1}$. The intensity of this component is found to be constant,
within the  errors, during  this observation and also when comparing with previous observation
($<I_N>=4.5\times 10^{-5}$ photons cm$^{-2}$ s$^{-1}$, \citealp{Balestra}; $I_N=6.5\pm 2.7\times 10^{-5}$
photons cm$^{-2}$ s$^{-1}$, \citealp{Weaver97}). The constancy in flux of this line together with the  limits
on the width obtained with \chandra\ are suggestive of an origin  from distant matter such as the putative
torus. Indeed  the   upper limit on the FWHM corresponds to a distance from the central black hole  greater
then 10$^4$R$_g$. 

 The presence of two Compton-thick  X-ray reprocessors   responsible for the two components of the iron line, suggested since
the \asca\ observation \citep{Weaver97} is confirmed with this deep \xmm\ observation and with the deep Suzaku observation
\citep{Reeves07}. The geometry  inferred for \sorg\  with this latter observation is discussed in detail in \citet{Reeves07}.
To summarize,  one plausible scenario is that we are seeing \sorg\ through the Compton thin  edge of the  putative torus, which
is Compton thick at the plane of the accretion disk. This is in agreement with the  inclination
($i=41^{+29}_{-12}$$^\circ$)\footnote{Modeling  the combined  Suzaku  and \xmm\ spectra, \citet{Reeves07} found  an inclination
of $i=53^{+7}_{-9}$ $^\circ$  which is in agreement within the errors with the value obtained from the analysis of the \xmm\ 
spectrum.} of the accretion disk derived by  modeling the broad line component with a  relativistic line  profile.  The
column density of $1.5\times 10^{22}$ cm$^{-2}$  measured using low energy cut-off  is thus associated with the  thinner
absorbing  material, e.g.   encountered viewing through the edge of the torus. We found no evidence of variability of the
column density of this absorber within this long observation, and also no strong variability is found when comparing the 
column densities measured with the previous observations performed with BeppoSAX \citep{Risaliti02a}, \asca\ \citep{Weaver97},
\xmm\ and \chandra\  \citep{Dewangan03, Balestra}.   This result implies that the absorber is probably  far from the central
black hole and   there is no evidence  that this absorber is  clumpy as suggested for other Seyfert 2s \citep{Risaliti02b}.   

This deep \xmm\  observation confirms  the presence of a relativistically
broadened  iron K$\alpha$ line;  the width derived from modeling this component with a  Gaussian profile
corresponds to a  FWHM $\sim 40000$ km s$^{-1}$ and   is  suggestive of  an origin in the accretion disk.
The profile of this component is nearly symmetric and  can  be  modeled, equally well with a broad Gaussian or a 
relativistic profile; in the latter case the derived inner radius is about 20--40$R_g$.  Since the advent
of \xmm\ and \chandra\, one of the most debated issues in the study of the broad iron K$\alpha$ lines has
been the  fraction of AGNs which clearly show the presence of a broad component \citep{Nandra06,
Guainazzi06}. Several authors have  discussed  the robustness of the detection in some
objects of  a broad iron line  (i.e. NGC~3516, \citealp{Turner05}; NGC~3783, \citealp{Reeves04}).  This
controversy  emerged when  observations  characterized by  high photon statistics showed  the ambiguity 
of modeling the iron K$\alpha$ line when complex absorption is present \citep{Pounds03, Pounds04}.  Indeed, a
high  column density warm absorber  can produce  curvature in the spectrum at the energy of the iron
K$\alpha$  line that mimics the profile of a relativistically broadened emission line. The detection
of the absorption feature at $\sim$7.8 keV shows that a high column density variable absorber  (a high
velocity, highly ionized outflow) is also present   in MCG--5-23-16; this  could in principle give rise  to
ambiguity in the interpretation of the broad component.  However, in the case of \sorg\,   the 
availability of a simultaneous observation with Suzaku above 10 keV  allowed us to tightly  constrain the
underlying X-ray continuum and to rule out the interpretation of the broad line as due to unmodeled
complex absorption. Indeed the residuals left  at the energy of the iron K$\alpha$, when  we take into
account the  amount of reflection detected with Suzaku,   cannot be explained  by the effect of 
complex absorption. Furthermore, once the  absorber responsible for the  feature detected at $\sim 7.8$
keV  is accounted for,  either in the time-averaged spectrum or  in the third segment of the \xmm\
observation,  a  broad line is still required by the data with a similar EW and FWHM. Note that the
ionization parameter of the absorber is required to be high and does not introduce additional spectral
curvature below 6 keV, which hence does not impact the broad iron line modeling.    

The remaining open
questions on the origin of the broad line in \sorg\ are the relatively large inner radius derived for the
accretion disk, and  its lack of variability.   The former can be explained  with several scenarios:  the
disk could be   truncated  or missing below 20R$_g$,   or  the inner part of the accretion disk could be 
so highly ionized  that the iron is fully ionized. However it worth noting that, as shown by \citet{Reeves07},   assuming a flat emissivity ($q=2$) a inner radius  of about 6R$_g$ (in the case of a Schwarzschild
black hole) cannot be statistically ruled out.

The second open issue  is the lack of variability of
the   iron emission line  both on short and long  time-scales. Indeed the flux of the broad component
is found to be consistent with being constant when using  the   short-term time-resolved spectroscopy  
performed  within this deep observation, and when comparing our result with the  long-term  flux history
presented in \citet{Balestra}. The strength of the  broad component  appears to be lower  during this
observation with respect to the  value reported since the first \asca\ observation (EW$\sim 200$ eV; \citealp{Weaver97}),
however when we take into account the larger errors on the early \asca\ measurements we cannot exclude the line being constant.\\  In
\sorg\ this lack of variability of the iron emission line is not so striking  as in MCG-6-30-15
(\citealp{Miniutti, Vaughan04}), due to the
low   level of variability of the intrinsic continuum (30-40\% compared to a factor of 2-3 in case  of 
MCG-6-30-15).

\subsection{The blue-shifted absorption line:  a possible high velocity, ionized variable outflow}   
Perhaps the most interesting result of this long \xmm\ observation has been the
discovery of a possibly variable absorption line from  ionized iron. The feature
appears to be transient with a  time scale of about 20 ksec and it is detected at
an  observed energy of about $7.66$ keV (corresponding to a rest-frame energy of
$7.72$ keV). As shown the most plausible association of this feature  is with
K-shell absorption from H-like iron, which is blue-shifted  by  $\sim 0.1c$. 
Indeed modeling this absorption feature with {\sc xstar}   \citep{Bautista01}
requires a column density of about $8\times 10^{22}$ cm$^{-2}$  and a high
ionization parameter (log$\xi=3.7\pm 0.3$ erg cm s$^{-1}$) which implies that the
absorption is  due to  a blueshifted   1 $\rightarrow$ 2 transition  of
{Fe\,\textsc{xxvi}} ($E=6.97$ keV). The velocity of the absorbing material is
found to be $v=(0.09\pm 0.01)c $. 

In the last few years,  red-  and blue-shifted
absorption lines associated with  the presence of  highly ionized  gas  in- 
and/or out-flowing at relativistic velocities  have been reported for several AGN
(E1821+643 \citealp{Yaqoob05}; Mkn~509 \citealp{Dadina05}; NGC~3516
\citealp{Nandra99, Turner05};  Mrk~335, \citealp{ Longinotti07}). These absorption
lines are found   both in Seyfert galaxies (NGC~3783, \citealp{Reeves04};  IC4392a,
\citealp{Markowitz06}; NGC~1365, \citealp{Risaliti05}; Ark~564,
\citealp{Papadakis07}) as well as in  quasars (PG~1211+143, \citealp{Pounds03};
PDS~456, \citealp{Reeves03}) and BAL QSOs  (APM~08279+5255, \citealp{Chartas02};
PG~1115+080,  \citealp{Chartas03}). These absorption systems can  also  be 
variable  on different time scales,  in  their ionization state   and column
density, with the most extreme cases being NGC~1365 \citep{Risaliti05} and 
Mkn~509 \citep{Dadina05}.   From the  analysis of the  different intervals in
which we split  the observation we can infer that it is unlikely that the 
variability of the absorber in \sorg\ is due to a change in the ionization state
of the  outflowing material, otherwise we would detect signatures of this
absorber  in all  the spectral slices. A more likely  scenario is a change in
column density  of this  absorber.  

The present data  suggest  we are seeing a   transient absorber, which   could be associated
with a   cloud which sporadically obscures the central source. This ``cloud''   could be 
the    signature of a clumpy absorber located close to the central X-ray source or    of
matter ejected sporadically. Different models have been proposed to explain the powerful
outflows detected  with the recent \xmm\ and \chandra\ observations; in particular 
transient red and blue shifted absorption lines are predicted in several theoretical models
of failed disk winds \citep{Proga00, King03}  or an aborted jet \citep{Ghisellini04}. The
picture emerging is that these features can provide a direct probe of the  dynamics and 
kinematics of the  innermost central regions of AGNs.  For  the absorber detected in \sorg\
the data suggest that this feature appears when the source reaches a relative maximum in
the   intrinsic 2--10 keV flux.  However, monitoring the spectrum on longer timescales would
be required to determine whether there is a statistically firm correlation between the
presence of the absorber and the source brightness and determine  if any duty cicle is present. Thus both a scenario
where  a clumpy  absorber  or a variable or failed outflow or jet  are at present possible.
Finally, it is worth noticing that the rapid variability of this absorption feature is
indicative of a compact  absorber and rules out a possible $z=0$ origin, e.g. due to  the warm
intergalactic medium (WHIM) or a  local hot bubble  as  claimed along the line of sight to  other  AGN
\citep{Mckernan04,Mckernan05}. \\ 

Before deriving an estimate of
the   location, mass and energetics associated with the absorber   we performed a
consistency check   for the $N_{\rm H}$   value measured with the {\sc xstar} model and the 
EW ($\sim 50$ eV) of the absorption line measured with the Gaussian component.
Following the curve of growth  for  H-like iron (see Fig.~4  of \citealp{Risaliti05})
we inferred that the  detected EW  requires a turbulence velocity greater than
500 km a$^{-1}$; a lower turbulence velocity would imply a Compton thick absorber and a
similar EW of the H-like K$\beta$ line (due to saturation of the K$\alpha$ line),
which is not detected ($|EW|< $15 eV at the 90\% confidence level).  On the other
hand a   turbulence velocity  greater than 3000 km s$^{-1}$  would produce a broad
absorption feature which would be  resolved even at  the EPIC-pn resolution at 8
keV ($\sim 170$ eV). We therefore conclude that the observed EW and line width
are in broad agreement with the {\sc xstar} estimate of a column density of about
$10^{23}$ cm$^{-2}$, for a turbulence velocity $\sigma=1000$ km s$^{-1}$. Using this 
value for the column density  we can now estimate the  maximum distance of this
cloud or blob from the central black hole  using the relation between the
ionization parameter, the  density of the absorber and the illuminating continuum
luminosity: $L/\xi=nR^2$; where $L$ is the intrinsic 2--100 keV  X-ray luminosity
($5.4\times 10^{43} $ erg s$^{-1}$).  Assuming then that the thickness of the cloud
$\Delta R=N_H/n$ is less than the distance R,  we  find $R< 10 ^{17}$ cm.  A
lower limit for the distance of the absorber,  assuming it is in the form of an
outflow,  can be obtained equating the velocity of the absorbing material to the
escape velocity at a given  radius R from the central black hole; the derived
distance is then $R\simgt 100 R_g$.  A constraint on the size of this cloud  can
be placed assuming 20 ksec  as the characteristic variability timescale, when our
line of sight intercepts the absorbing cloud; this gives us $\Delta R \sim 6\times
10^{13}$ cm, which corresponds to $\sim 10\; R_g$   for a black hole mass of
$5\times 10^{7} M_{\odot}$ \citep{Wandel86}. We can then infer a 
density of $\sim 10^{9}$ cm$^{-3}$ and assuming a spherical geometry for the
cloud, a mass of about $10^{28}$ g.   These one order of magnitude  estimates  
for the  mass and  velocity correspond to a kinetic energy $E_{\rm{kin}}\sim
5\times10^{46}$ erg and using the 20 ksec as a characteristic timescale to a
power  of $\sim 2.5\times 10^{42}$ erg s$^{-1}$. This value corresponds to $\sim$10\% of
the 2--10 keV X-ray luminosity and is thus is agreement with a radiation driven
wind model \citep{Proga04} or with an aborted jet \citep{Ghisellini04}.    \\ 

In conclusion,  this deep \xmm\ observation revealed that the soft
X-ray emission of \sorg\ can be ascribed to  material photoionized by the AGN,
likely to be located outside the sub-pc scale of the absorber and  perhaps
coincident with the NLR. We confirm the presence of a iron K$\alpha$ emission
line complex composed by a   narrow    and a broad relativistic component. The
inclination  derived  from the diskline profile  ($i\sim40^\circ$)  is in
agreement    with the orientationally based  Unification Scheme of AGN
\citep{Antonucci},  the X-ray classification  of \sorg\ as Compton thin Seyfert
2 (i.e. intermediate between a type 1  AGN and a Compton thick  type 2) and the
optical classification as a Seyfert 1.9.  Finally we detected a sporadic Fe K
absorption feature which could be a signature of a variable high velocity
outflow. This detection adds one more example to the   increasing sample of AGN
where relativistic outflows have been revealed in the X-ray band. The growing
evidence of high velocity outflows in AGN indicates that they may play an
important role in the energetics of AGN central engines.    

\acknowledgments 
We would like to thank the anonymous referee for his/her useful comments that have improved this paper.
This paper has made use of observations obtained with XMM-Newton, an ESA science mission with instruments and contributions directly funded by ESA Member States and the USA (NASA).
Support for this work was provided by the National Aeronautics and Space Administration through Chandra Award Number  GO5-6146Z issued by the Chandra X-ray Observatory Center, which is operated by the Smithsonian Astrophysical Observatory for and on behalf of the National Aeronautics Space Administration under contract NAS8-03060. 

\clearpage

\newpage
\begin{figure}
\begin{center}
\rotatebox{-90}{
\epsscale{0.5}
\plotone{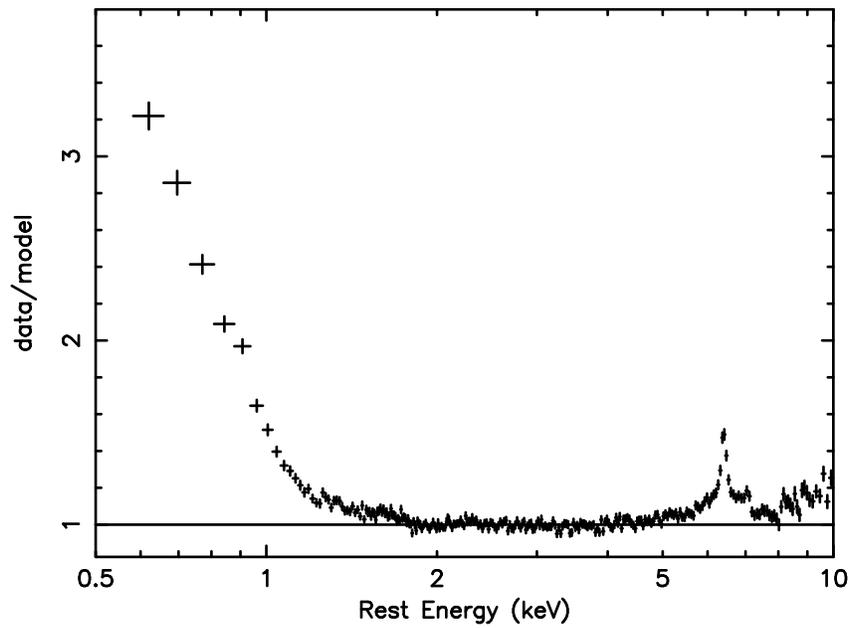}}
\caption[fig1.eps]{Ratio  of the \xmm\ pn data and the model when fitting an absorbed  and redshifted power law  over the 2-10 keV band ignoring the iron emission
line  energy range ($5.5-7.5$ keV). The photon index of the  power law has been fixed to 1.8. Two different  residuals are  clearly present: a soft X-ray excess 
below $\sim 2$  keV; and the Fe complex at $\sim 6.4$  keV.
 \label{fig:ratiopl}}
\end{center}
\end{figure}

\begin{figure}
\begin{center}
\rotatebox{-90}{
\epsscale{0.5}
\plotone{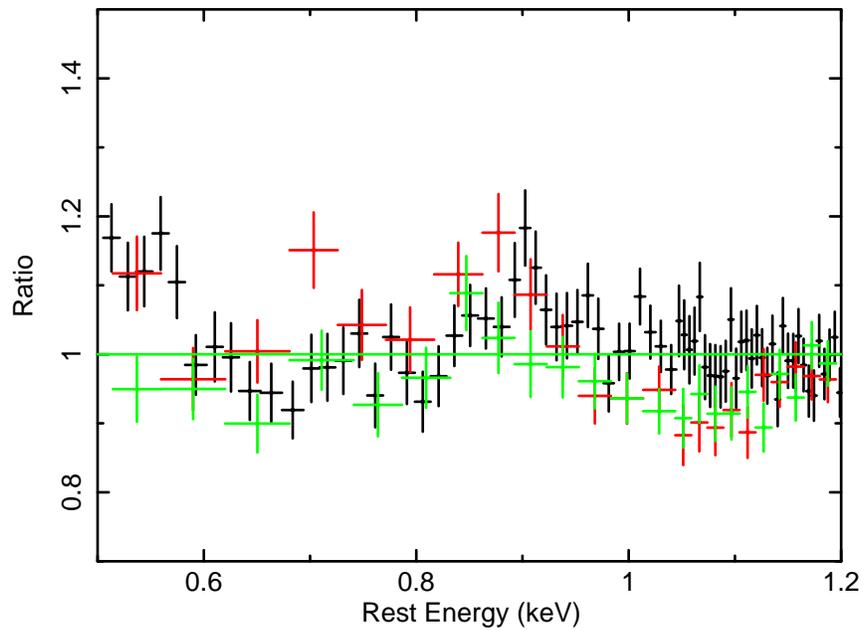}}
 \figcaption[fig2.eps]{
The ratio between the \xmm\ pn (black points), MOS1 (red), and MOS2  (green) data and the the model, plotted between 0.5-1.5 keV. The model consists of
the baseline absorbed power-law described in the text and an additional soft ($\Gamma=3$) power law, absorbed only by a Galactic column. An emission line is clearly
detected at 0.9 keV.
\label{fig:pnsoft}}
\end{center}
\end{figure}

\clearpage
\begin{figure}
\begin{center}
\includegraphics*[width={0.3\columnwidth}, angle=-90]{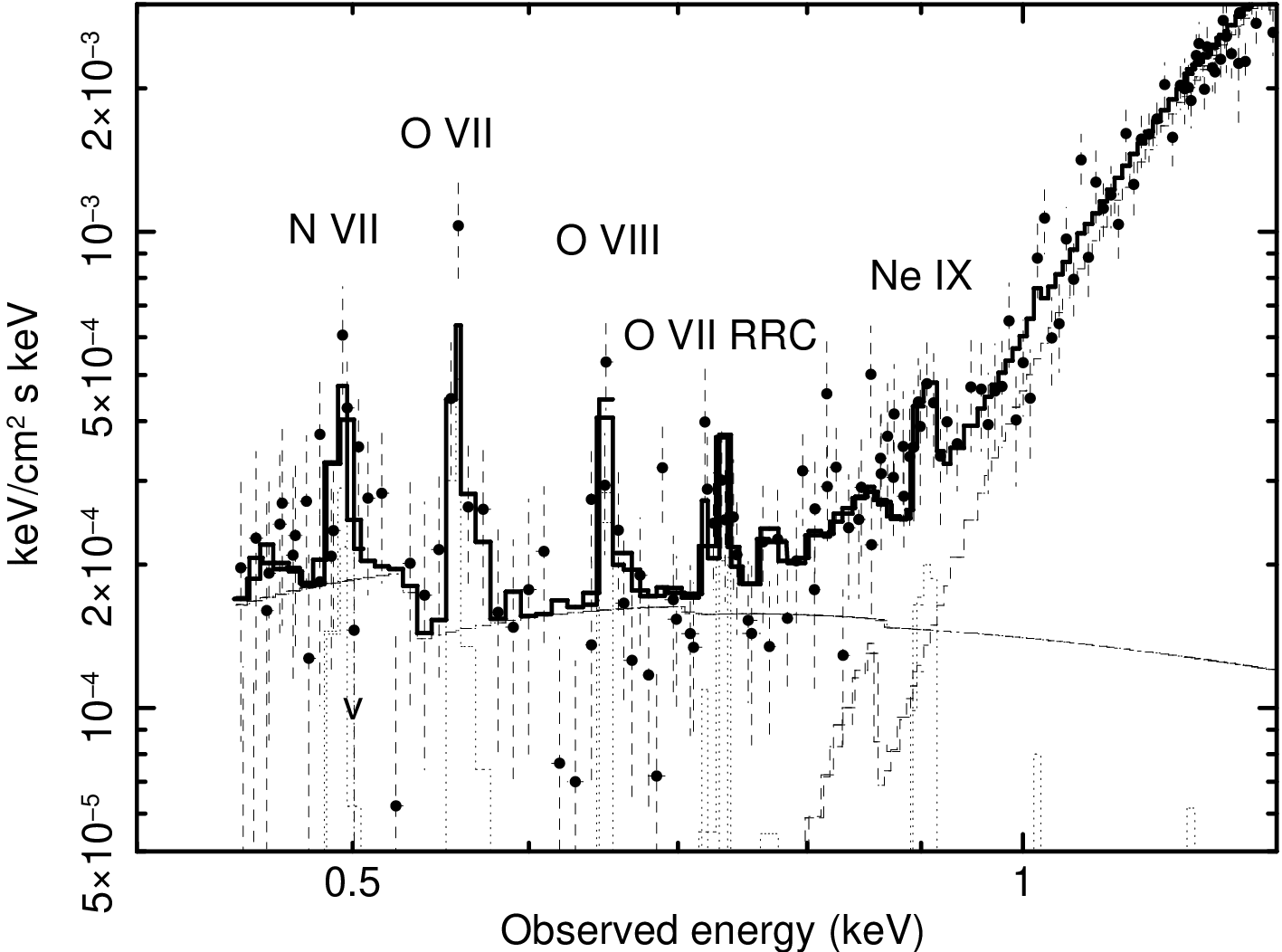}
\hfil
\includegraphics*[width={0.3\columnwidth}, angle=-90]{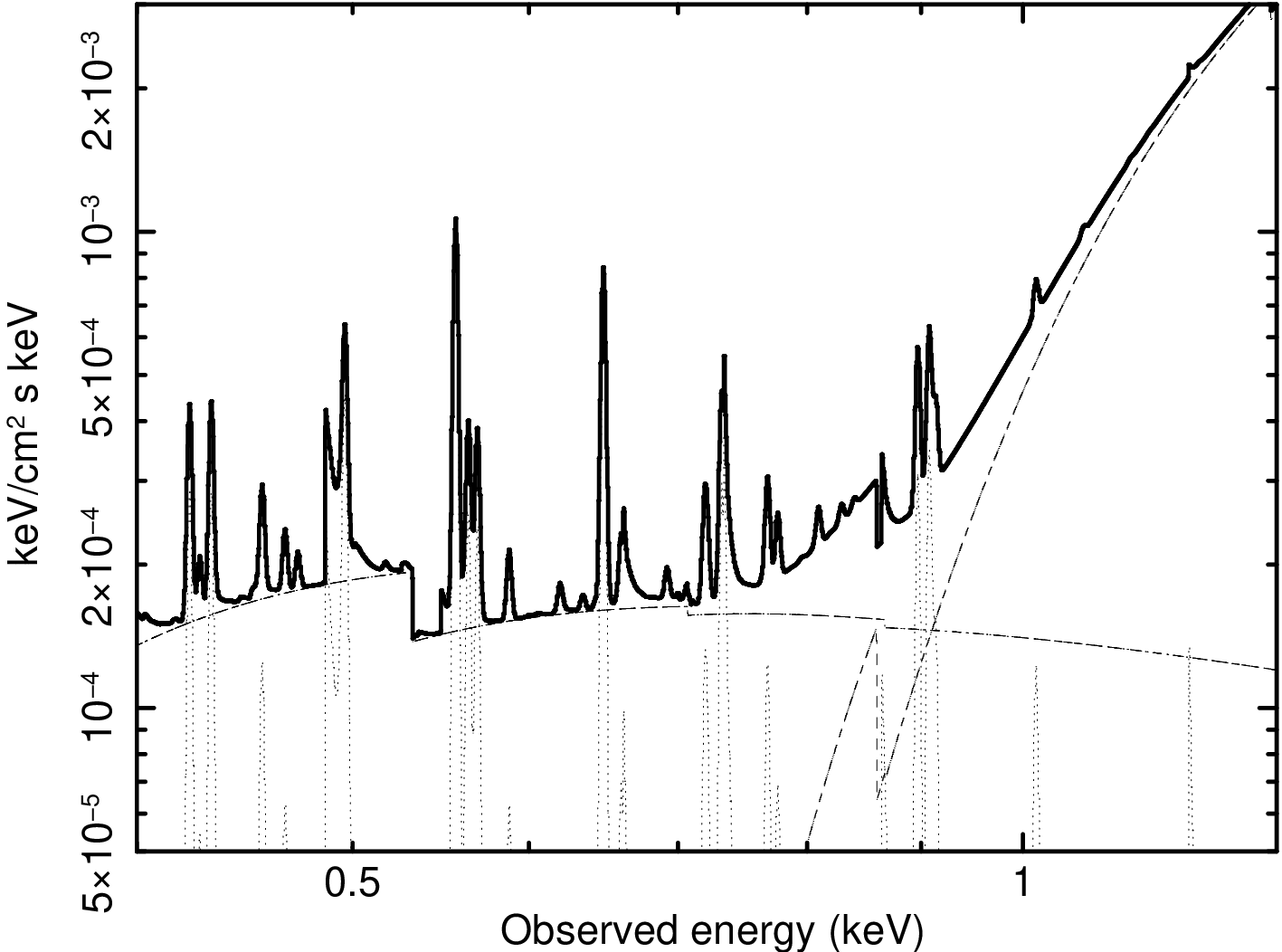}
\caption[fig3.eps]{
Left panel: \xmm\ RGS1 and RGS2  spectra in the 0.4$-$1.3 keV  energy range. The underlying AGN continuum model is composed of
an absorbed power law   plus a scattered  soft power-law component ($\Gamma=1.82$). Several emission lines 
 are clearly  detected. The possible identifications  of the 5 brightest lines are also shown. Above $\sim 0.9$ keV 
 a steep rise of the continuum is evident, due to the emergence of the  absorbed power-law component. Right panel: best fit   photoionized plasma model (in the
 0.4$-$1.3 keV)  which includes a    {\sc xstar}  component  with  an ionization parameter $log \xi=1.29^{+0.17}_{-0.13}$ (see Section 3.2).

\label{fig:rgs.ps}}
\end{center}
\end{figure}

\begin{figure}
\begin{center}
\rotatebox{-90}{
\epsscale{0.5}
\plotone{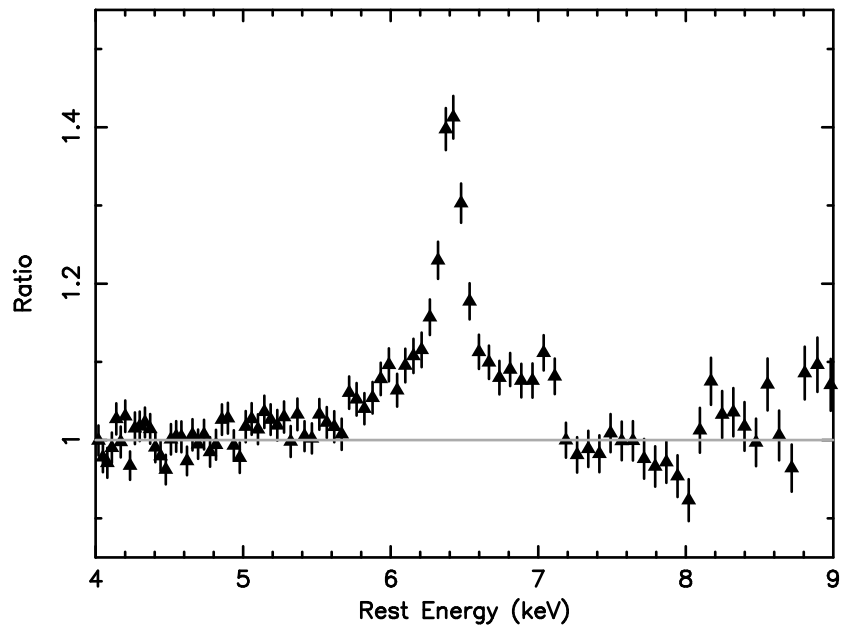}}
\caption[fig4.eps]{
Ratio   between the pn data and the absorbed power-law model ($\Gamma=1.65$) showing the  iron line profile. The data clearly show a narrow core at 6.4 keV, red and blue wings  extending from 5.7  keV to 7  keV and a narrow emission feature at $\sim$7.05 keV, which is due to  Fe K$\beta$. A sharp drop is also present at $\sim 7.1$ keV due to  presence of reflection.
 \label{fig:pnline.ps}}
\end{center}
\end{figure}
\clearpage

\begin{figure}
\begin{center}
\rotatebox{-90}{
\epsscale{0.5}
\plotone{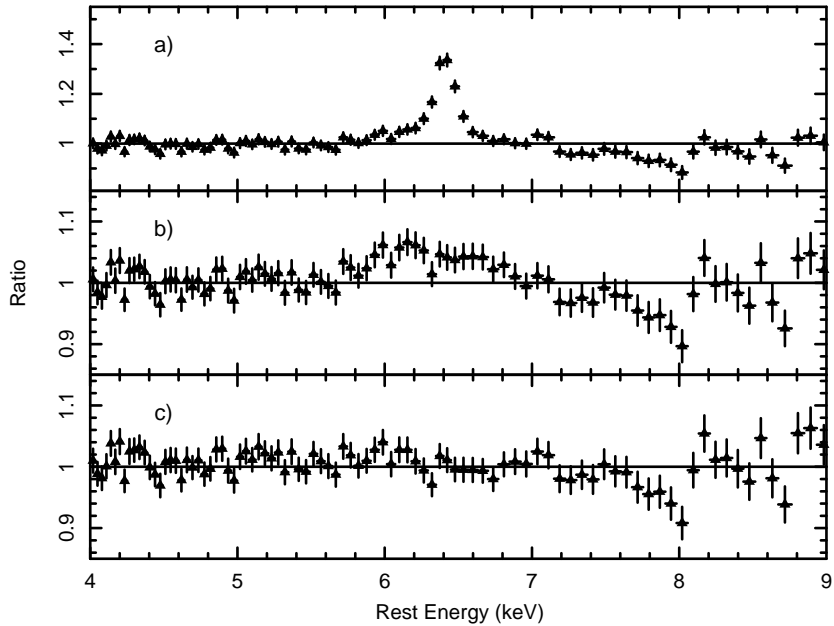}
}
\caption[fig5.eps]{
Data/model residuals  for the pn data at the Fe K band. Panel  (a)  shows  the data/model ratio when the  underlying
continuum is composed of an absorbed power law plus a reflection component ($R=1.1$). Panel (b) shows the  residuals left  when  the narrow Fe K$\alpha$
and Fe K$\beta$ lines are added to the model. An excess between 5.8  keV and 7 keV is still present. Panel (c) shows the residuals when the  broad
component of the Fe K$\alpha$  line is fitted with a diskline model. The overall fit is now good and only a weak absorption feature is
left at 8 keV.
\label{fig:pnpanel.ps}}
\end{center}
\end{figure}

\begin{figure}
\begin{center}
\rotatebox{-90}{
\epsscale{0.5}
\plotone{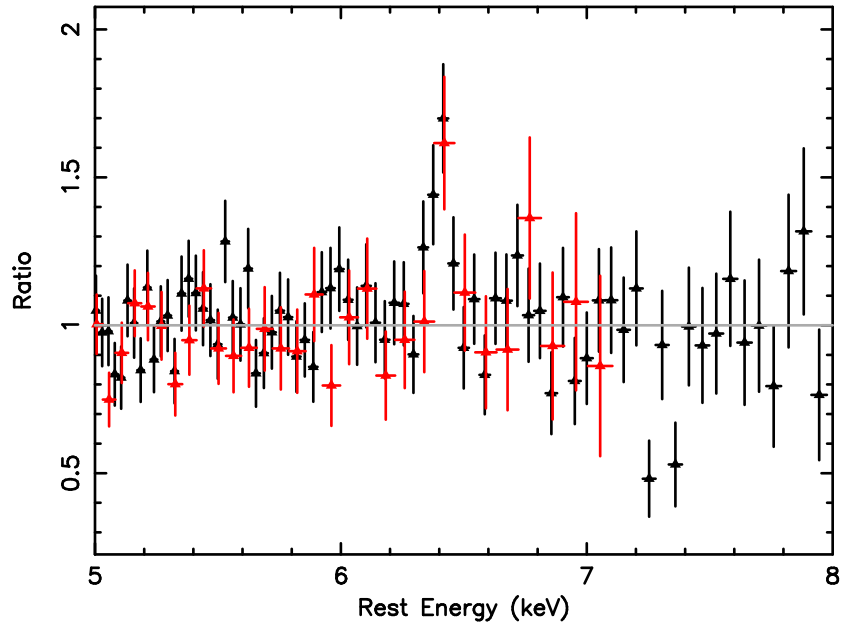}
}
\caption[fig6.eps]{Data/model ratio for the \chandra\ HETG.  The AGN continuum is composed of an absorbed power law  plus a reflection
component  ($\Gamma=1.82$; $R=1.1$)  A strong narrow core at 6.4 keV is observed. When modeled with a single Gaussian, the line  is found to
have $EW\sim 80$ eV and a    width     $\sigma=32^{+19}_{-16}$ eV. 
\label{fig:chandra.ps}}
\end{center}
\end{figure}

\begin{figure}
\begin{center}
\rotatebox{-90}{
\epsscale{0.5}
\plotone{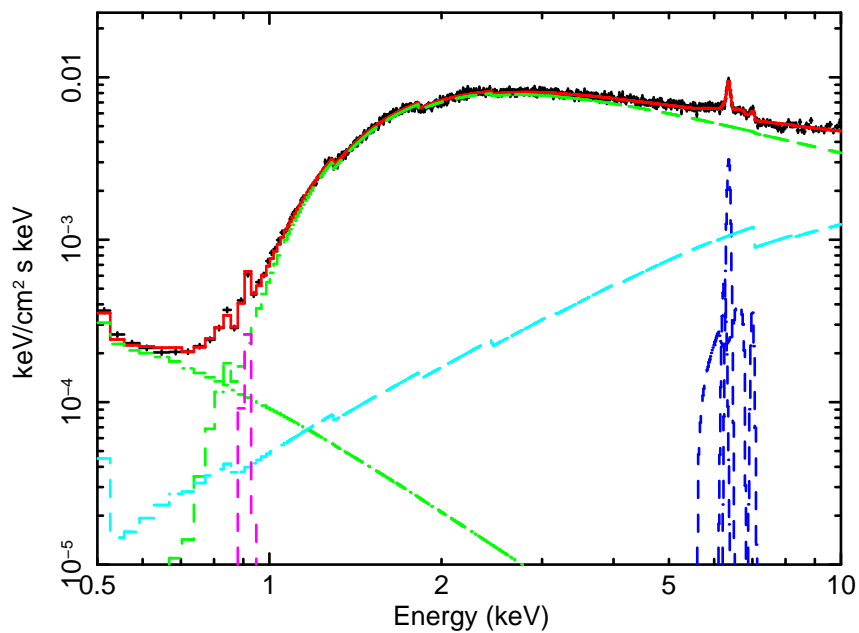}
}
\caption[fig7.eps]{\xmm\  spectrum of \sorg. The black points denote the EPIC-pn data. The red line is the total best fit model. The AGN continuum model is
composed of:  an absorbed and a scattered power-law components (green lines online version) plus a reflection component (light blue line online
version). The
Fe K line complex  (blue line online version) is composed of  narrow and  broad Fe K$\alpha$ lines and  a  narrow K$\beta$ line.  A soft X-ray
emission line at $\sim0.9$ keV is also shown (magenta line online version).
  \label{fig:tot.ps}}
\end{center}
\end{figure}

\begin{figure}
\begin{center}
\rotatebox{-90}{
\epsscale{0.5}
\plotone{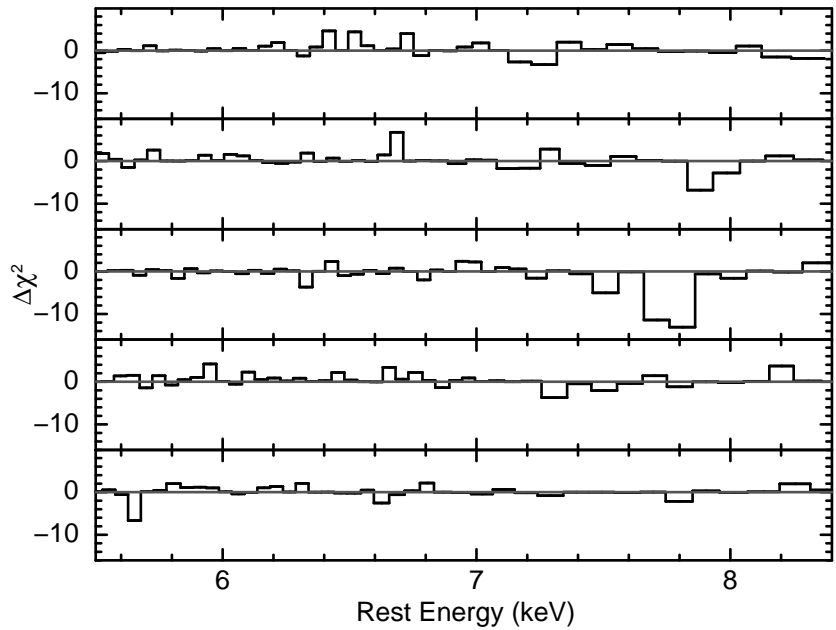}}
\caption[fig8.eps]{Contribution to the $\chi^2$ in the 5.5--8.4 keV  for the five pn  spectra extracted with a time bin of  20
ksec. The Fe K$\alpha$  line has been  parameterized with two Gaussian lines; all the  parameters of the model except the
primary power-law   normalization have been kept tied together. The only strong deviation in  $\Delta\chi^2$ is  present  in the third spectrum at 7.7 keV.}
\label{fig:chi_intervals.ps}
\end{center}
\end{figure}

\begin{figure}
\begin{center}
\rotatebox{-90}{
\epsscale{0.7}
\plotone{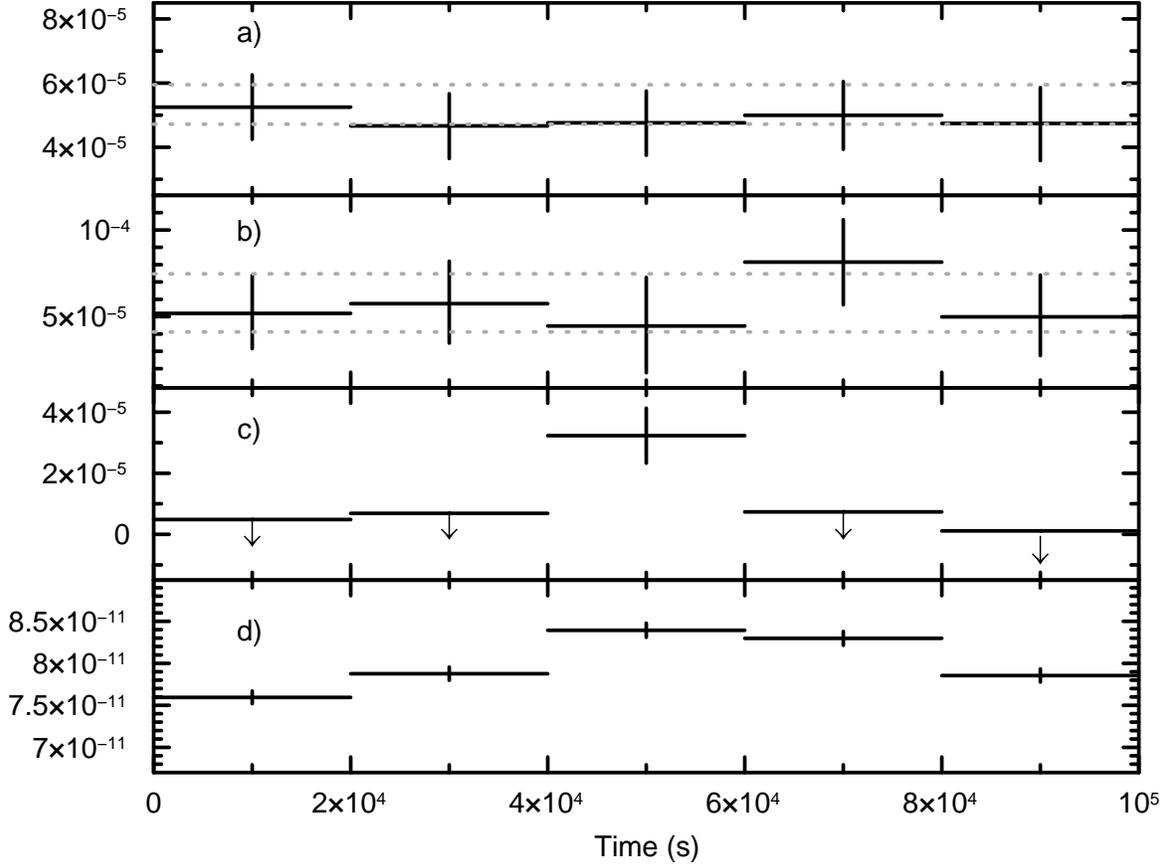}}
\caption[fig9.eps]{Time  resolved spectral analysis. Panel (a): Fe 6.4 keV narrow core intensity (in unit of
photons cm$^{-2}$ s$^{-1}$) versus the time intervals. The dashed lines correspond  to the 90\% confidence level of
the  normalization of the  narrow core measured in the average spectrum.  Panel (b):  same as panel a)   for
the broad component. Panel (c): Absolute  intensity of the absorption feature (in unit of photons cm$^{-2}$
s$^{-1}$), the line  energy has been fixed to  the best fit value found for the third interval (7.71 keV).  Panel
(d): \sorg\ 2--10 keV flux (in erg cm$^{-2}$ s$^{-1}$) . Error bars and upper limits are at the  90\% confidence
level.}
\label{fig:comparison.ps}
\end{center}
\end{figure}

\begin{figure}
\begin{center}
\rotatebox{-90}{
\epsscale{0.5}
\plotone{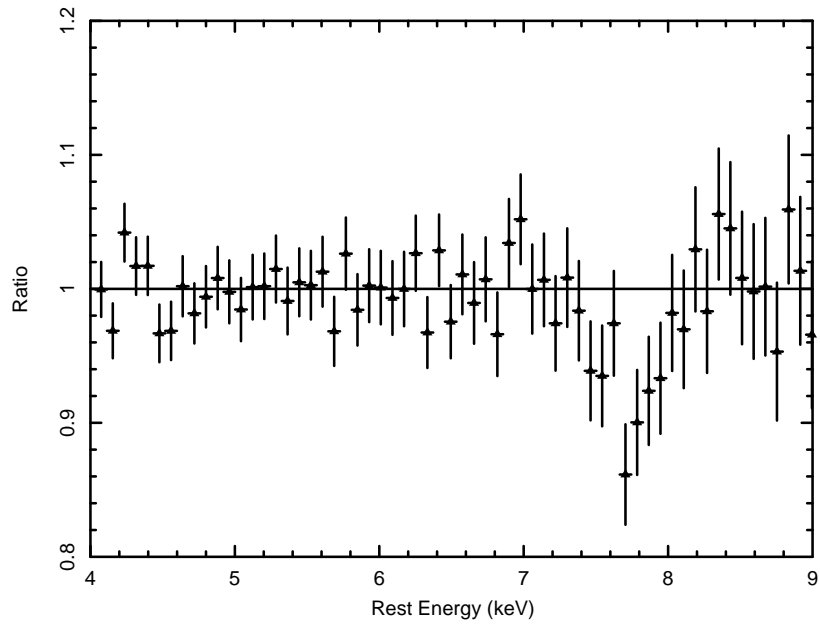}
}
\caption[fig10.eps]{Data/model ratio of the pn spectrum of the third 20 ksec interval, with the data binned with a constant energy bin of 80 eV. A deep absorption feature is visible at 7.7 keV (see text for details).
\label{fig:energubin.ps}}
\end{center}

\end{figure}

\begin{figure}
\begin{center}
\rotatebox{-90}{
\epsscale{0.5}
\plotone{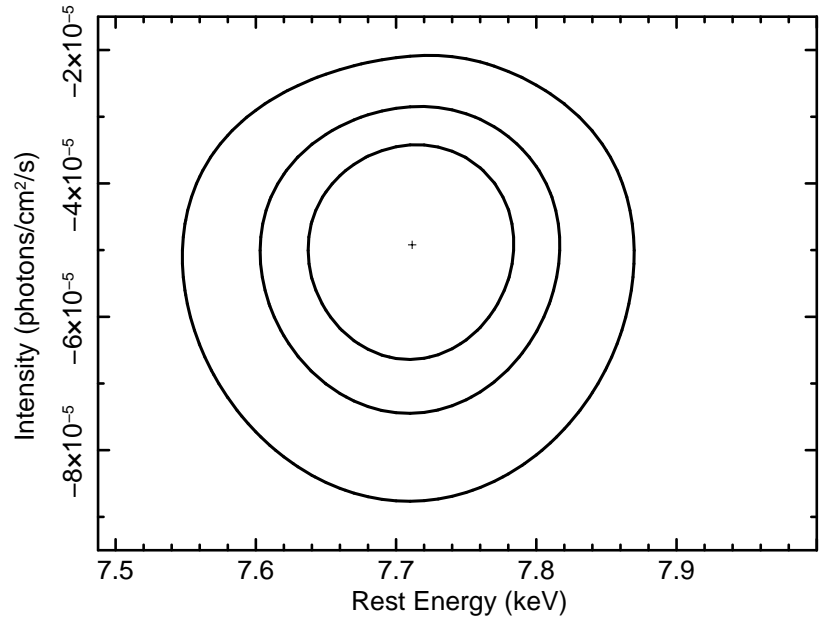}
}
\caption[fig11.eps]{ Derived  99\%, 90\% and 68\% confidence contours of the absorption line intensity  vs. the observer-frame energy.
\label{fig:contorni.ps}}
\end{center}
\end{figure}

\begin{figure}
\begin{center}
\rotatebox{-90}{
\epsscale{0.3}
\plotone{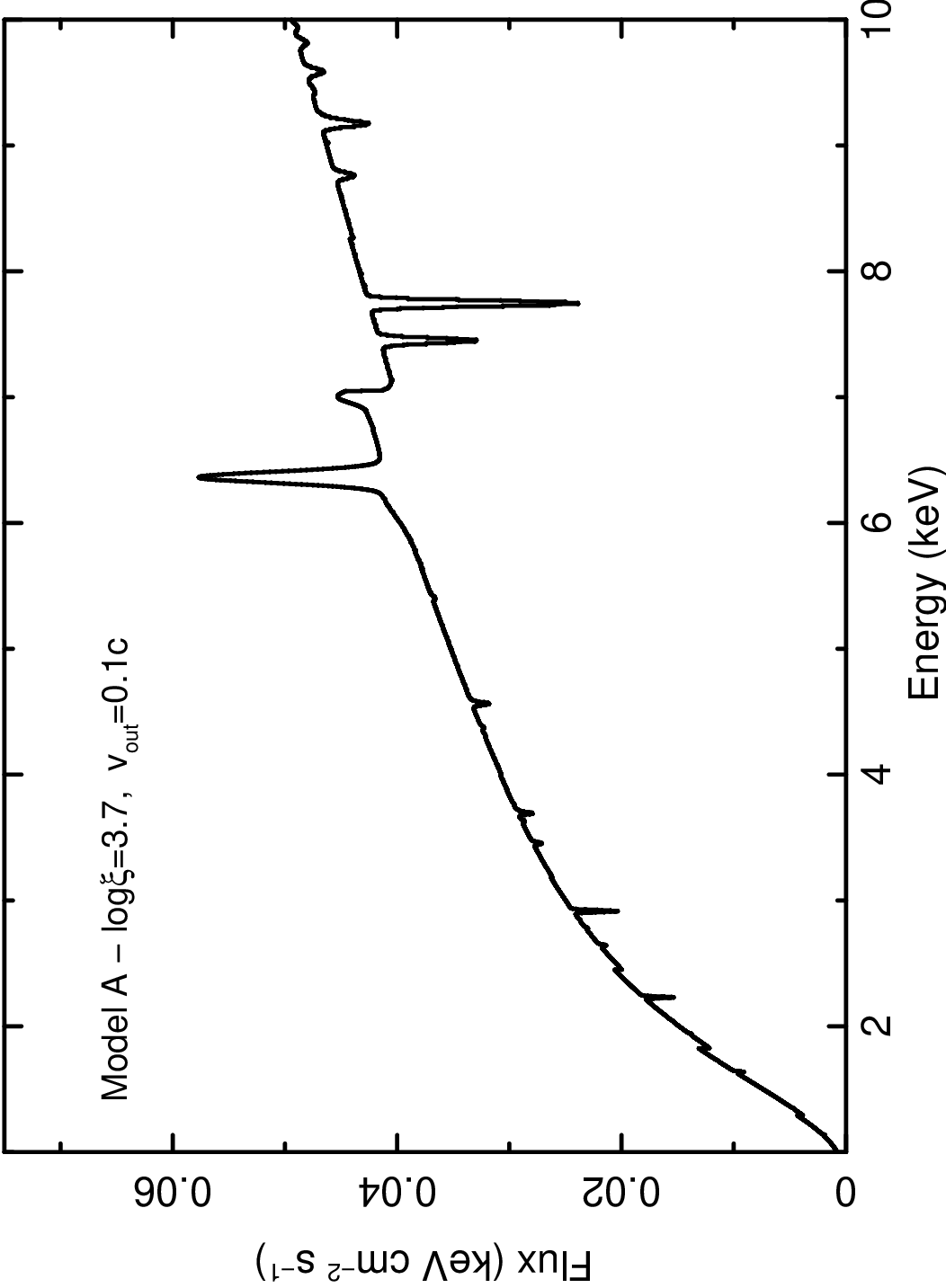}
\plotone{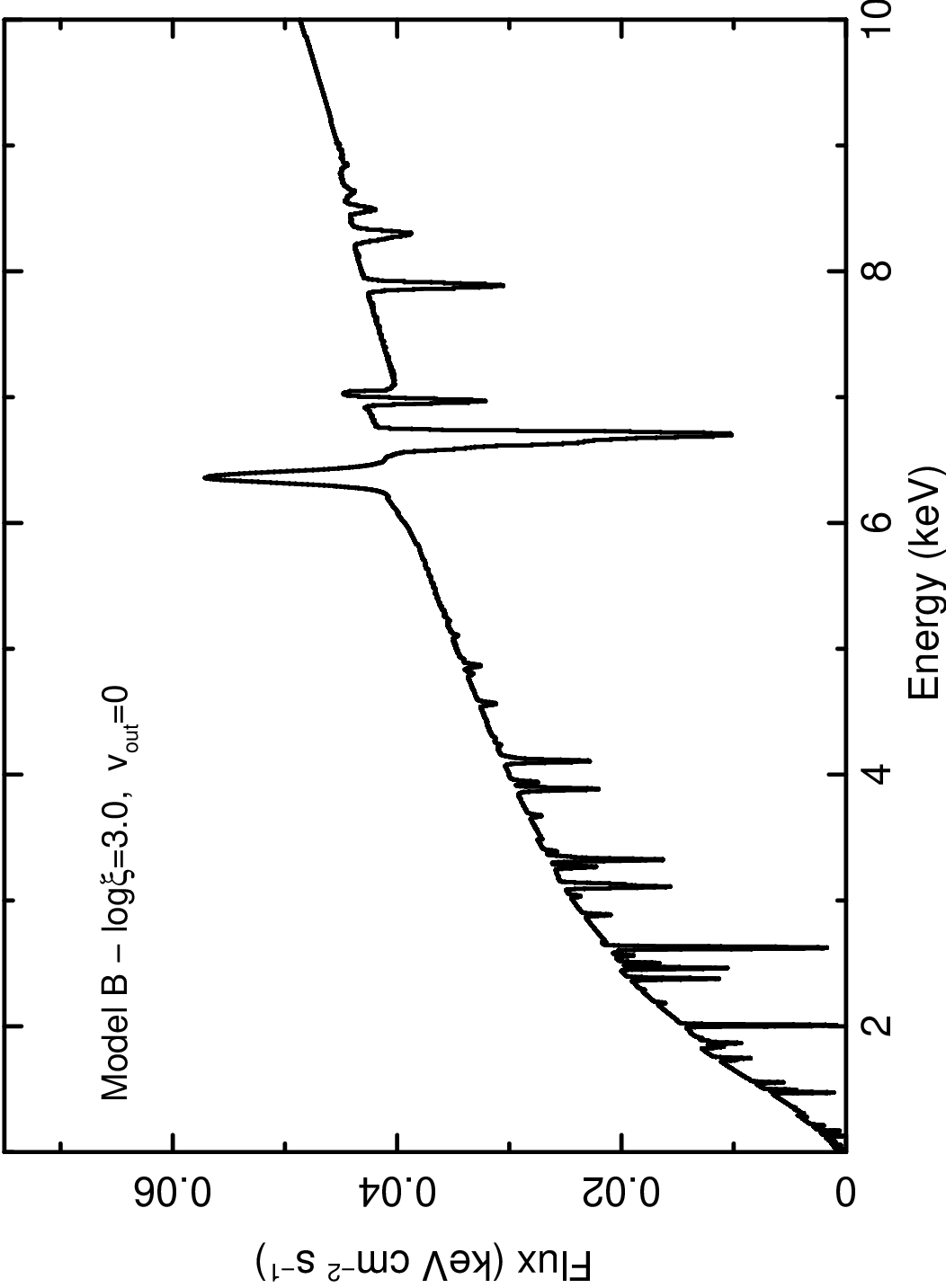}
 \plotone{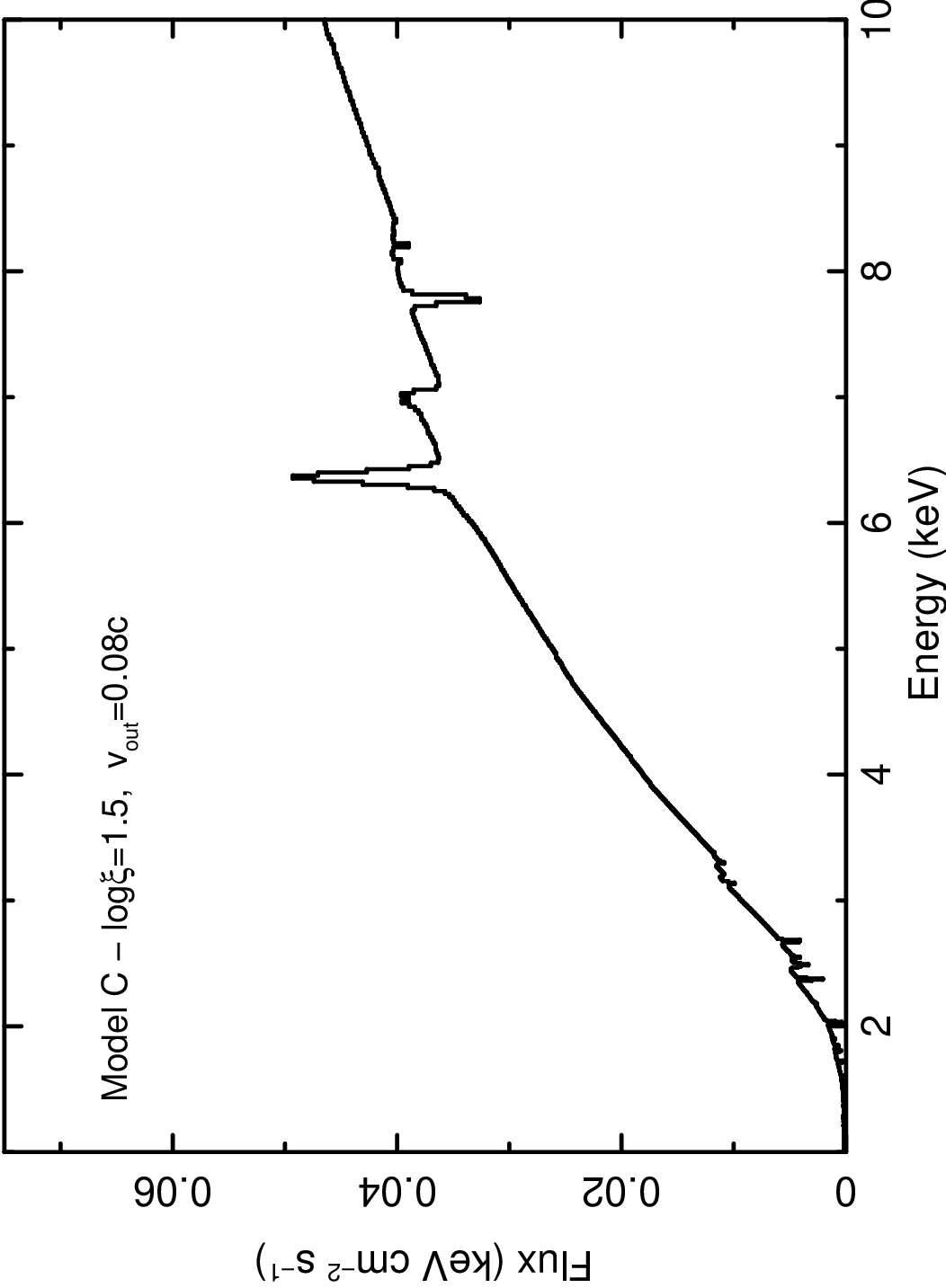}
}
\caption[fig12.eps]{Example of high ionization warm absorber models. Panel(a) shows the best fit model: an
ionization state characterized by $log\xi=3.7^{+0.2}_{-0.3}$ erg cm s$^{-1}$  and a  column density 
of $ \sim 8\times 10^{22}$ cm $^{-2}$   with an outflow velocity of $0.09\pm 0.01\,c$. The {Fe\,\textsc{xxvi}} Ly$\alpha$
absorption feature shifted to 7.7 keV. fits well the observed dip at this energy. Panel (b): the  absorption feature at 7.7 keV
can be accounted for  by assuming  a column density of
$N_{\rm H}=10^{23}$\,cm$^{-2}$ and  an ionization parameter of $log\xi=3.0$ erg cm s$^{-1}$.   However, the model predicts stronger absorption due to the {Fe \,\textsc {xxv}} 1s-2p
absorption line at 6.7 keV, which is not detected.  The model in panel (c) shows that  lower ionization  K$\beta$ absorption line without strong K$\alpha$ from the same species, can be obtained assuming  a column density of $N_{\rm
H}=10^{23}$\,cm$^{-2}$, an ionization parameter of $log\xi=1.5$ erg cm
s$^{-1}$.  In this scenario a blueshift of
$\sim0.08c$ would still  be required to model the absorption line at
7.7\,keV in the spectrum and the column density 
required to model the EW of the K$\beta$ absorption feature is
$N_{\rm H} >> 10^{23}$\,cm$^{-2}$; which would introduce too much continuum  absorption below 6 keV, inconsistent with the
pn data. } 
\label{fig:xstar.ps}
\end{center}
\end{figure}
 \clearpage
  
\begin{deluxetable}{cccccc}
    \tabletypesize{\scriptsize}
  \tablecaption{Log of the observations and exposure times.}\label{tab:log_observ}
\tablehead{
\colhead{Mission} & \colhead{Instrument} & \colhead{T$_{\rm(total)}$ (ks)} & \colhead{T$_{\rm(net)}$ (ks)}
& \colhead{T$_{\rm {START}}$} &
\colhead{T$_{\rm {STOP}}$}
}
  
  \startdata

              &  &  &    &&\\

 \xmm\ & PN& 131.5   &96.2	&08/12/2005 21:11:28&10/12/2005 09:17:08\\
 \xmm\ & MOS1& 131.5 &101.6	&08/12/2005 20:42:51&10/12/2005 09:16:48\\
 \xmm\ & MOS2& 131.5 &102.8	&08/12/2005 20:43:51&10/12/2005 09:16:53\\
 \xmm\ & RGS1& 131.6 &97.2	&08/12/2005 20:41:37&10/12/2005 09:18:03\\
 \xmm\ & RGS2& 131.6 &97.2	&08/12/2005 20:41:42&10/12/2005 09:18:03\\
 \chandra\ & ACIS-S HETG& 30&-&	 08/12/2005 17:41:30&09/12/2005 02:33:59\\
 \chandra\ & ACIS-S HETG& 20&-&	 09/12/2005 20:52:11&10/12/2005 03:00:10\\

  \enddata
\tablecomments{For \xmm\ the exposure values   reported are total and net exposure time after  filtering for high-background time intervals.}
\end{deluxetable}

\begin{deluxetable}{cccccc}
\tablecaption{\xmm\ RGS1 and RGS2: best fit emission lines required  fitting with a photoionization model.}
\tablewidth{0pt}

\tablehead{

 \colhead{Energy}     & \colhead{Flux} &\colhead{ID}       & \colhead{EW}& \colhead{$\Delta C$}& \colhead{E$_{\rm {Lab}}$}\\
   \colhead{(keV)} & \colhead{($10^{-5}$ ph cm$^{-2}$ s$^{-1}$)}&\colhead{}&\colhead{(eV)}&&\colhead{(keV)}\\
 
    \colhead{(1)}  &  \colhead{(2)}  &  \colhead{(3)}   &  \colhead{(4)}  & \colhead{(5)} & \colhead{(6)} \\
}
   \startdata    
0.499$^{+0.001}_{-0.001}$& 0.68$^{+0.39}_{-0.29}$& {N\,\textsc{vii}} Ly$\alpha$&$10\pm5$& 14.3& 0.500\\   
   &   &   &     & &\\

      &   &   &     & &0.561 (f)\\

0.564$^{+0.002}_{-0.002}$& 2.30$^{+1.42}_{-1.05}$& {O\,\textsc{vii}} He$\alpha$&$46.5^{+28.3}_{-20.7}$&22.5&0.569 (i)\\
      &   &   &     & &0.574 (r)\\
   &   &   &     & &\\
0.653$^{+0.001}_{-0.001}$& 0.61$^{+0.37}_{-0.29}$& {O\,\textsc{viii}} Ly$\alpha$  &14.7$^{+9.0}_{-7.0}$ &15.4&0.654 \\
   &   &   &     & &\\
0.732$^{+0.007}_{-0.007}$& 0.54$^{+0.40}_{-0.33}$& {O\,\textsc{vii}} RRC &15.7$^{+11.0}_{-10.0}$&7.0&$>0.739$\\
    &   &   &     & &\\
 &   &   &     & &0.905 (f)\\

0.903$^{+0.020}_{-0.020}$& 0.42$^{+0.26}_{-0.28}$& {Ne\,\textsc{ix}} He$\alpha$&$15.9\pm10.0$&10.4& 0.915 (i)\\
  &   &   &     & &0.922 (r)\\

  \enddata

\tablecomments{The model of underlying AGN continuum has been parametrized according to the
best fit model obtained with the analysis of the pn spectrum (section 3). The $\Gamma$ of the soft power law  is tied to the 
hard power-law component and fixed to the best fit value ($\Gamma=1.82$). The energy of the lines are   quoted in the rest frame. Fluxes and possible identifications
are reported in col.~(2) and (3) respectively. The EWs are reported in col.~(4) and they are  calculated against the soft power
law component. In col.~(5) the improvement of fit is
shown using  the $C$-statistic; the value for the model with no lines is $C$=455.7 for 358 PHA bins. In col.~(6) we report the theoretical value for the transition energies.}

\end{deluxetable}

\begin{deluxetable}{lcccc}
\tablecaption{Results of the fit  for the mean spectrum and the low and high flux states.}
\tablecolumns{4}
\tablewidth{0pc}

\tablehead{\colhead{ } & \colhead{Parameter} & \colhead{Mean}   & \colhead{High}    & \colhead{Low} }
\startdata
Continuum	& $\Gamma$ &$1.82\pm 0.01$ &$1.84\pm 0.01$ &$1.81\pm0.01$ \\
		&N$_{\rm H}^a$&$1.49\pm 0.01$ &$1.50\pm 0.02$ &$1.49\pm 0.02$\\
		&Flux$^b$  & 8.16& 8.84&7.39\\
 && & &\\
\hline 
 && & &\\

Narrow  Gaussian &E   & $6.42\pm0.01$&$6.41\pm0.02$ &$6.40\pm0.02$\\

 &EW  &$61^{+7}_{-7}$ &48$^{+12}_{-13}$ &81$^{+15}_{-15}$\\
&N$^c$ &$5.4^{+0.6}_{-0.6}$ &4.6$^{+1.3}_{-1.2}$ &6.5$^{+1.2}_{-1.2}$\\

& & & &\\
Broad Gaussian&E   &6.4 $^f$ &6.4 $^f$ &6.4 $^f$\\
  &$\sigma$  &$0.35\pm 0.1$ & 0.35$^{+0.15}_{-0.13}$&0.37$^{+0.22}_{-0.13}$\\
&EW   &   64$^{+18}_{-16}$ & 61$^{+17}_{-23}$ &$75^{+32}_{-32}$\\
&N$^c$ &$5.9\pm 1.5$ &5.9$^{+2.7}_{-2.6}$ &6.3$^{+2.9}_{-2.6}$\\
& & & &\\

  Diskline &  R$_{in}$     &48$^{+62}_{-20}$ &... &...\\  
&$i$&41$^{+29}_{-12}$&...&...\\
&EW &53$^{+14}_{-13}$ & ...  & ...\\

&N$^c$ &$4.6\pm 1.2$ &...&...\\
\enddata

\tablecomments{The line  energies are expressed in units of  keV, while theirwidths  $\sigma$ and EWs are in  eV. The disk 
radial emissivity has been fixed to $q=3$.}
\tablenotetext{a}{Column density in units of $10^{22}$ cm$^{-2}$}
\tablenotetext{b}{2--10 keV  flux in units of 10$^{-11}$ erg cm$^{-2}$ s$^{-1}$}
\tablenotetext{c}{Normalization  of the Fe line in unit of 10$^{-5}$ photons cm$^{-2}$ s$^{-1}$}
\tablenotetext{f}{Indicates that the parameter  has been kept fixed.}
\end{deluxetable}

\begin{deluxetable}{lcccc}
\tablecaption{Best fit parameters for the absorpion feauture detected at 7.7 keV for the  EPIC pn and MOS cameras and SUZAKU.\label{tab:bas_feat}}
\tablecolumns{4}
\tablewidth{0pc}

\tablehead{\colhead{ Mission-Instrument } & \colhead{Energy} & \colhead{Flux}   & \colhead{$|EW|$}\\    
\colhead{}   &\colhead{(keV) } & \colhead{($10^{-5}$ph cm$^{-2}$ s$^{-1}$)} &  \colhead{(eV)}  \\  
}

   \startdata

 EPIC-pn$^a$&$7.9\pm0.1$&$-1.9\pm0.6$&$33^{+9}_{-11}$\\
                &  &  &\\
EPIC-pn$^b$&$7.7\pm0.1$&$-3.2\pm0.9$&$52\pm 15$\\
                &  &  &\\
EPIC-MOS$^a$&$7.4\pm0.2$&$-1.1\pm1.0$&$16\pm 14$\\
                &  &  &\\
EPIC-MOS$^b$&$7.4\pm0.2$&$-2.2\pm1.8$&$27\pm 23$\\
                &  &  &\\
Suzaku-XIS0-1-2-3$^a$&$7.8\pm0.1$&$-1.8\pm0.9$&$30\pm 15$\\
                &  &  &\\
Suzaku-XIS0-1-2-3$^{b,c}$&$7.65\pm0.4$&$-3.5\pm3.0$&$52\pm 44$\\

\enddata

\tablecomments{The parameters of the absorption line are derived adding  an inverted gaussian component to the best fit model.  }
\tablenotetext{a}{Parameters refers to  mean spectrum}
\tablenotetext{b}{Parameters are derived  for the 20 ksec time  interval where the line is detected.}
\tablenotetext{c}{The net exposure of this Suzaku  spectum  is only $\sim 6$ ksec}
\end{deluxetable}

\end{document}